\documentclass[twocolumn]{aastex631}

\newcommand{\mathbfit}[1]{\textbf{\textit{#1}}}

\usepackage{newtxtext,newtxmath}

\begin{document}

\title{Modeling YSO Jets in 3D I: Highly Variable Asymmetric Magnetic Pressure-Driven Jets in the Polar Cavity from Toroidal Fields Generated by Inner Disk Accretion}

\author[0000-0003-2929-1502]{Yisheng Tu}
\affiliation{Astronomy Department, University of Virginia, Charlottesville, VA 22904, USA}
\affiliation{Virginia Institute of Theoretical Astronomy, University of Virginia, Charlottesville, VA 22904, USA}

\author{Zhi-Yun Li}
\affiliation{Astronomy Department, University of Virginia, Charlottesville, VA 22904, USA}
\affiliation{Virginia Institute of Theoretical Astronomy, University of Virginia, Charlottesville, VA 22904, USA}

\author{Zhaohuan Zhu}
\affiliation{Department of Physics and Astronomy, University of Nevada, Las Vegas, NV, 89154-4002, USA}
\affiliation{Nevada Center for Astrophysics, University of Nevada, Las Vegas, 4505 S. Maryland Pkwy, Las Vegas, NV, 89154, USA}

\author{Chun-Yen Hsu}
\affiliation{Astronomy Department, University of Virginia, Charlottesville, VA 22904, USA}
\affiliation{Virginia Institute of Theoretical Astronomy, University of Virginia, Charlottesville, VA 22904, USA}

\author{Xiao Hu}
\affiliation{Astronomy Department, University of Virginia, Charlottesville, VA 22904, USA}
\affiliation{Department of Astronomy, University of Florida, Gainesville, FL 32608, USA}



\begin{abstract}
Jets and outflows are commonly observed in young stellar objects (YSOs), yet their origins remain debated. Using 3D non-ideal magnetohydrodynamic (MHD) simulations of a circumstellar disk threaded by a large-scale open poloidal magnetic field, we identify three components in the disk-driven outflow: (1) a fast, collimated jet, (2) a less collimated, slower laminar disk wind, and (3) a magneto-rotational instability (MRI)-active turbulent disk wind that separates the former two. At high altitudes, the MRI-active wind merges with the laminar disk wind, leaving only the jet and disk wind as distinct components. The jet is powered by a novel mechanism in the star formation context: a lightly mass-loaded outflow driven by toroidal magnetic pressure in the low-density polar funnel near the system’s rotation axis. A geometric analysis of the magnetic field structure confirms that magnetic tension does not contribute to the outflow acceleration, with magnetic pressure acting as the dominant driver. While the outflow in our model shares similarities with the magneto-centrifugal model-such as angular momentum extraction from the accreting disk-centrifugal forces play a negligible role in jet acceleration. In particular, the flow near the jet base does not satisfy the conditions for magneto-centrifugal wind launching. Additionally, the jet in our simulation exhibits strong spatial and temporal variability. These differences challenge the applicability of rotation-outflow velocity relations derived from steady-state, axisymmetric magneto-centrifugal jet models for estimating the jet’s launching radius. For the slower disk wind, vertical motion is driven by toroidal magnetic pressure, while centrifugal forces widen the wind’s opening angle.

\end{abstract}

\keywords{Accretion (14) --- Circumstellar disks (235) --- Jets (870) --- Magnetic fields(994) --- Magnetohydrodynamical simulations(1966) --- Young stellar objects (1834)}


\section{Introduction}
Outflows from young stellar objects (YSOs) play a crucial role in the process of star formation. Observationally, they bridge the dynamics at small scales near the star and disk to larger-scale phenomena, providing indirect evidence of the physical processes occurring close to the star and along the disk surface. Outflows from YSOs are generally categorized into two components: the jet and the disk wind. Jets are highly collimated, fast outflows that can extend hundreds of au or even parsecs from the YSO \citep[][]{McGroarty2004, Ray2021, Whelan2024, Chauhan2024}, while disk winds are slower and less collimated, observable primarily closer to the star \citep[][]{Klaassen2013, Bjerkeli2016, Launhardt2023}. 
Both components play essential roles in star formation by removing angular momentum from the disk (and potentially the star), thereby driving accretion onto the star and facilitating disk evolution \citep[][]{Bouvier2007, Frank2014, Pascucci2023}. Despite their importance, the mechanisms driving jets and disk winds, as well as whether they share a common origin, remain areas of active research \citep[][]{Pudritz2007, Shang2007, McKinney2007, Bai2016, Jacquemin2019, Banzatti2022, Lopez2024, Zimniak2024}.

Numerous studies have explored the driving mechanisms of outflows from YSOs over the past decades, with a growing consensus that they are most likely magnetically driven \citep[][]{Pudritz2007, Shang2007, Ray2021}. While the presence of magnetic fields in the disk is expected \citep[][]{Wurster2018, Vlemmings2019, Galametz2020, Tu2024a, Ohashi2025}, the exact process of utilizing these fields to launch outflows remains a topic of active debate. 
The leading model for jets in YSOs is the class of magneto-centrifugal models, including the disk-wind model \citep[][]{Blandford1982, Pudritz1983} and the X-wind model \citep[][]{Shu1994, Shu2000, Ferreira2000}. In this class of models, rigid magnetic field lines efficiently transport angular momentum from the disk to the gas above. If the field lines are sufficiently inclined, the centrifugal force can overcome gravity along the lines, leading to strong gas acceleration, like ``beads on a wire." This accelerated gas is subsequently collimated into a jet, either by magnetic tension forces or through external confinement. To maintain the rigid magnetic field lines necessary for efficient centrifugal gas acceleration, only a small amount of gas can be loaded onto these lines, resulting in a ``lightly loaded" wind \citep[][]{Bai2013, Hawley2015}. This model effectively explains angular momentum transport and jet launching, and is particularly well developed under the assumptions of steady-state and axisymmetry \citep[e.g.][]{Ferreira2006}, which are often used to interpret observational data \citep[e.g.][]{Lopez2024}.

Another class of models suggests that jets are primarily powered by Poynting flux \citep[][]{Shibata1986, Lovelace2001, Lovelace2009}, with toroidal magnetic energy dominating over kinetic energy in the jet acceleration region. A related model is the so-called ``magnetic tower model,'' where field lines form loop-like structures stacked atop one another \citep[][]{Contopoulos1995, Lynden1996, Lynden2003}. The gas acceleration in this scenario is more gradual compared to the classic lightly-loaded magneto-centrifugal model since the field is more heavily loaded and the acceleration results from the expansion driven by magnetic pressure within the tower. 

Both classes of models have undergone extensive investigations using numerical simulations. The magneto-centrifugal models are investigated with 2D axisymmetric \citep[e.g.][]{Casse2002, Kato2002, Fendt2011, Fendt2013, Tzeferacos2009, Stepanovs2016, Ramsey2019, Mattia2020a, Mattia2020b, Jannaud2023, Tu2024c, Zanni2013, Zanni2009} and 3D \citep[e.g.][]{Anderson2006, Zhu2018, Mishra2020, Jacquemin-Ide2021, Takasao2022} simulations, typically with ideal MHD or a prescribed (turbulence) diffusivity that focuses on the outer wind-launching region. The Poynting-flux/magnetic tower model was investigated with a prescribed disk property \citep[][]{Ustyugova2000, Huarte-Espinosa2012}. A 3D model with a more realistic treatment of non-ideal MHD effect based on thermal ionization focusing on the innermost disk that distinguishes between the disk active zone and dead zone \citep[][]{Flock2016} is still lacking. Furthermore, a detailed force analysis in the jet-launching region is needed to determine the jet-launching mechanism. These will be the focus of this paper.

A primary objective is to identify the dominant jet-launching mechanism in YSOs and determine whether it aligns more closely with the classic magneto-centrifugal model or the magnetic tower model. This study extends the 2D axisymmetric model of \citet[][]{Tu2024c} to 3D to investigate the jet-launching mechanism in greater detail. In the 2D model, the ``avalanche accretion stream'' along the disk surface was identified as a key driver of the jet, transferring angular momentum from the accretion stream to the low-density gas above via open magnetic field lines. Here, we aim to determine whether this feature persists in 3D and retains its role in jet launching. To overcome the limitations of the previous approach, we transition from spherical-polar to Cartesian coordinates, resolving challenges posed by the polar and radial boundaries and allowing the jet to evolve naturally around the polar axis. The stellar magnetosphere is excluded in this study to isolate the role of the disk’s magnetic field and its associated disk accretion in jet launching; the magnetosphere will be included in a subsequent investigation. 

This paper is organized as follows: we describe our model setup in sec.~\ref{sec:methods}. Sec.~\ref{sec:result} shows an overview of the simulation result and a description of the outflow component in our model. The forces driving each outflow component are also discussed in sec.~\ref{sec:result}. The jet-launching mechanism is discussed in detail in sec.~\ref{sec:jet_details}, and a comparison to the expectations of steady-state jet solution and the magneto-centrifugal model is discussed in sec.~\ref{sec:compare_to_steady_state}. Sec.~\ref{sec:discussion} focuses on the implications of our jet-launching model on observation and interpretations of the jet-launching mechanism. We conclude in sec.~\ref{sec:conclusion}.

\section{Methods}
\label{sec:methods}
In a jet-launching inner disk around a young stellar object (YSO), the following set of equations governs the disk accretion and jet-launching processes: 

\begin{equation}
    \frac{\partial\rho}{\partial t} + \nabla\cdot(\rho\mathbfit{v}) = 0,
\end{equation}
\begin{equation}
    \rho\frac{\partial \mathbfit{v}}{\partial t} + \rho(\mathbfit{v}\cdot\nabla)\mathbfit{v} = -\nabla P + \frac{1}{c}\mathbfit{J}\times\mathbfit{B} - \rho\mathbfit{g},
    \label{equ:mhd momentum equation}
\end{equation}
\begin{equation}
    \frac{\partial \mathbfit{B}}{\partial t} = \nabla\times(\mathbfit{v}\times\mathbfit{B}) - \frac{4\pi}{c}\nabla\times(\eta_O\mathbfit{J}),
    \label{equ:mhd induction}
\end{equation}
where $\mathbfit{J} = (c/4\pi)\nabla\times\mathbfit{B}$ is the current density, and $\eta_O$ the Ohmic dissipation coefficient. Other symbols have their usual meanings. 

We use the ATHENA++ code \citep{Stone2020} with static mesh refinement (SMR) to solve the governing equations in Cartesian coordinates. The simulation domain is cubic, extending from $-10\ \mathrm{au}$ to $10\ \mathrm{au}$ in all three directions. The base resolution is $64^3$, and we use 7 levels of mesh refinement to focus on the jet-launching inner disk region, yielding a smallest cell size of 0.0024~au. The finest level (level 7) extends from -0.15~au to 0.15~au in the $\hat{x}$ and $\hat{y}$ directions and from -0.08~au to 0.08~au in the $\hat{z}$ direction, covering the magnetically active inner zone of the disk in the simulation. To better resolve the launched jet without significantly increasing computational cost, an additional refinement of level 6 is added, from -0.15~au to 0.15~au in $\hat{x}$ and $\hat{y}$ directions, and from -0.32~au to 0.32~au in $\hat{z}$ directions. The rest of the refinement is automatically generated, enforcing the criteria that adjacent mesh blocks must have refinement levels that differ by exactly one. This refinement strategy is chosen to optimize the resolution of the inner disk and the jet-launching region while maintaining a reasonable computational cost.

Since a major goal of this work is to extend the 2D (axisymmetric) simulations of \citet[][]{Tu2024c} to 3D, our initial setup of the problem will closely follow theirs. Specifically, the following power-law is adopted for the midplane gas density:
\begin{equation}
    \rho_\mathrm{mid}(r) = \rho_0\Big(\frac{r}{r_0}\Big)^p,
\end{equation}
where $\rho_0 = 10^{-9}\mathrm{\ g\ cm^{-3}}$; $r_0 = 6.3\times10^{-2}\ \mathrm{au}$ and $p = -1.35$. The midplane density within 0.06~au of the origin is gradually reduced to a floored density profile (see equation~\ref{equ:dfloor} below) to mimic the expected transition between the inner disk and circumstellar cavity caused presumably by the stellar magnetosphere, which is not included in this study to isolate the role of the disk magnetic field and its associated accretion in jet launching; the combined effects of the disk and stellar magnetic fields will be explored elsewhere. The disk density away from the midplane is calculated assuming vertical isothermal hydrostatic equilibrium with a dimensionless disk scale height of $h = H/R = 0.1$ (where $H$ is the scale height and $R$ is the cylindrical radius). We extrapolate the disk's vertical density profile into the atmosphere above and below the disk.  

To help with numerical stability, particularly in the polar regions where {the magnetic field is expected to be strong and gas is depleted due to rapid infall}, we implement a similar radially varying density floor as in \citet{Tu2024c}\footnote{ The inclusion of a stellar wind, as done by \citet[][]{Meskini2024}, may alleviate the need for the density floor in future simulations.}:
\begin{equation}
    \rho_\mathrm{floor}(r) = 
    \begin{cases}
      \rho_f ;             & r < r_f, \\
      \rho_f(r / r_f)^{-2}; & \text{otherwise},
    \end{cases}
    \label{equ:dfloor}
\end{equation}
where $\rho_f = 5.94\times10^{-17}$~g cm$^{-3}$ and $r_f = 0.025$~au. Both the disk and envelope are initially in Keplerian rotation. To speed up the development of potential instabilities such as current-driven instabilities in toroidal magnetic field-dominated parts of jets, such as the nonaxisymmetric helical kink m = 1 mode \citep[see, e.g.][]{McKinney2009}, a random perturbation with a magnitude of up to 1\% of the local velocity is added to the system. This is in addition to the grid-level numerical noise that can also break the axisymmetry of the initial conditions and seed potential instabilities.

The temperature in the disk (where ${\vert z/r\vert} < 0.1$) follows that of \citet[][]{Flock2016}:
\begin{equation}
    T_\mathrm{disk}(r) = T_i \frac{T_\odot}{\epsilon_T^{0.25}}\left(\frac{R_\odot}{2r}\right)^{1/2}
\end{equation}
where $T_\odot=5666$~K and $R_\odot = 6.943\times10^{10}$~cm are the solar surface temperature and radius respectively; $\epsilon_T = 1/3$. Following \citet{Tu2024c}, we set the parameter $T_i$ as 
\begin{equation}
    T_i = 
    \begin{cases}
        2 & r < 0.1\ \mathrm{au}, \\
        3 - \frac{r}{0.1\ \mathrm{au}} & 0.1\ \mathrm{au} \leq r < 0.2\ \mathrm{au}, \\
        1 & \mathrm{otherwise}.
    \end{cases}
\end{equation}
to increase the midplane temperature inside 0.2~au to mimic accretion heating and produce a well-defined magnetically active inner disk region inside $\sim 0.13$~au.
A temperature transition zone lies between the disk and the envelope, where the temperature is given by
\begin{equation}
    T_\mathrm{trans}(r) = (T_\mathrm{env} - T_\mathrm{disk})\frac{\vert z/r\vert - 0.1}{0.2} + T_\mathrm{disk}; \ \ 0.1\leq \Big\vert\frac{z}{r}\Big\vert \leq 0.3,
\end{equation}
where $T_\mathrm{env} = 3,000\ \mathrm{K}$ is the temperature of the envelope above the transition zone (where $|z/r| > 0.3$); { it is adopted primarily to ensure that the envelope remains well coupled to the magnetic field as generally expected, although a more self-consistent temperature treatment is desirable in future investigations.} For simplicity and to ensure numerical stability, we assume that the spatial distribution of the temperature remains the same throughout the simulation.

The initial magnetic field is calculated from a vector potential $\mathbfit{A} = \langle 0, 0, A_\phi \rangle $ following \citet{Wang2019}
\begin{equation}
    A_\phi = \frac{4}{3} \frac{A_{z, 0}(r/r_0)^{-0.25}}{[1+1/(\mu\tan\theta)^2]^{0.625}},
\end{equation}
where $\mu = 0.5$ determines how curved the initial field is. The scale for the magnetic field strength is chosen to be $A_{z, 0} = 0.009~\mathrm{G}~\mathrm{au}$ so the plasma-$\beta$ at $1~\mathrm{au}$ is about $10^5$.

\subsection{Sink Treatment}

Since our focus is on outflow driven by disk accretion, we will adopt a simplified treatment of the central star and ignore, for now, the stellar magnetosphere. We define a spherical ``sink" region of radius $r_\mathrm{fix} = 0.025~\mathrm{au}$ 
(the same as the radius of the plateau density floor $r_f$ in equ.~\ref{equ:dfloor}), where all hydrodynamic quantities are reset to their initial values at each step, but the magnetic field is allowed to evolve over time. To gauge the potential effects of stellar rotation on jet launching, we adopt two simple prescriptions for the velocity field in the sink region within $r_\mathrm{fix}$. In the ``\textsc{norot}'' model, the velocity is set to 0 in all three directions. In the ``\textsc{strot}'' model, the gas within $r_\mathrm{fix}$ is assumed to rotate as a solid body with an angular velocity equal to half the Keplerian rate at $r_\mathrm{fix}$. This translates to a stellar spin period of approximately 2.9 days, typical of those observed in YSOs \citep{Herbst2007}. 

Our sink treatment also includes a softening of the stellar gravity: 
\begin{equation}
    a_g(r) = 
    \begin{cases}
      -\frac{GM_\star\hat{\mathbfit{r}}}{r^2}\ \frac{(r - r_s)^2}{(r - r_s)^2 + r_s^2} ;             & r > r_s, \\
      0; & \text{otherwise},
    \end{cases}
\end{equation}
where $M_\star=1M_\odot$ is the stellar mass and $r_s = 0.005~\mathrm{au}$ is the softening radius, which resides well within the sink region of radius $r_\mathrm{fix}$. The gravitational acceleration is close to the unsoftened value (within $\sim 5\%$) outside the sink region ($r = r_\mathrm{fix}$). 

\subsection{Magnetic diffusivity}

Our simulation captures both the inner magnetically active disk and magnetically inactive dead zone at larger distances by including Ohmic dissipation. The coefficient $\eta_O$ is calculated by combining the contributions of two ionization sources: cosmic ray ionization and thermal ionization, following \citet[][]{Tu2024c}. It is given by \citep{Bai2011}
\begin{equation}
    \eta_O = \frac{c^2}{4\pi\sigma_O},
    \label{equ:etaO}
\end{equation}
where $c$ is the speed of light. The conductivity $\sigma_O$ is given by
\begin{equation}
    \sigma_{O} = \frac{ec}{B}\sum_j n_jZ_j\beta_j,
\end{equation}
with
\begin{equation}
    \beta_j = \frac{Z_j eB}{m_j c}\frac{1}{\gamma_j\rho},
\end{equation}
where $n_j, m_j, Z_j,$ and $ \gamma_j$ are the number density, mass, electric charge, and momentum transfer rate between the $j$th charged species and the neutrals, respectively. The parameter $\beta_j$ characterizes the relative importance between the Lorentz force and the neutral drag.  

The conductivity due to cosmic ray ionization is computed using two charged species: the ions and the electrons. The ion abundance is given by
\begin{equation}
    n_i = \frac{C\sqrt{\rho}}{m_i}
\end{equation}
where the coefficient $C = 9.5\times 10^{-18}~{\rm cm}^{-3/2}~{\rm g}^{1/2}$ is determined by a cosmic ray ionization rate of $10^{-20} \ \mathrm{s}^{-1}$. The value is chosen to account for the screening effect due to the disk column density and the ionization due to decaying radioactive atoms \citep{Umebayashi2009}. The momentum transfer coefficient between the ions and the neutral is $\gamma_\mathrm{i, cr} = 3.5\times10^{13}\ \mathrm{cm^3\ s^{-1}}$ \citep{Shu1992}. The electron abundance is assumed to be the same as the ion abundance. The electron momentum transfer coefficient is given by \citep{Bai2011}
\begin{equation}
    \gamma_e = 8.3\times10^{-10}\sqrt{T}\ \mathrm{cm^3\ s^{-1}},
\end{equation}
where $T$ is the temperature of the gas. 

The conductivity due to thermal ionization is computed using five charged species: potassium, calcium, magnesium, sodium, and electron. The total number density of each metal is obtained from its relative abundance to $H_2$ shown in Table~\ref{tab:ions}. The number density of each metal ion is calculated with the Saha equation, and the electron density is taken as the sum of the number densities of all metal ions. The momentum transfer rate between the metals and the neutral is given by \citep{Bai2011}
\begin{equation}
    \gamma_j = \frac{2\times10^{-9}}{m_n + m_j}\Big(\frac{m_H}{\mu}\Big)^{1/2}\ \mathrm{cm^3\ s^{-1}},
\end{equation}
where 
\begin{equation}
    \mu = \frac{m_j m_n}{m_j + m_n}
\end{equation} 
is the reduced mass in the collision between the $j$th metal ion and the neutral (with mass $m_n$).

The total conductivity is computed by summing the contributions of each component, and the diffusivity is calculated according to equ.~\ref{equ:etaO}.

\begin{table}
    \centering
    \begin{tabular}{c|c|c|c}
         Species & $m_s$ [$m_p$] & $n/n_\mathrm{H2}$ & $e_I$ [erg]\\
         \hline
         K & $39$ & $2.14\times10^{-9}$ & $6.95\times10^{-12}$ \\
         Ca & $40$ & $4.38\times10^{-8}$ & $9.76\times10^{-12}$ \\
         Na & $23$ & $3.48\times 10^{-8}$ & $8.2\times 10^{-12}$ \\
         Mg & $24$ & $7.96\times 10^{-7}$ & $1.2\times 10^{-11}$
    \end{tabular}
    \caption{Parameters for calculating the thermal ionization fraction with the Saha equation. $m_s$, the atomic mass of each element, is in the unit of proton mass $m_p$. The relative abundances with respect to H$_2$ are taken from \citet{Asplund2009}, assuming the gas phase metal in the protoplanetary disk is 1\% of the solar photosphere abundance.}
    \label{tab:ions}
\end{table}

\section{Result overview}
\label{sec:result}

\begin{figure}
    \centering
    \includegraphics[width=\columnwidth]{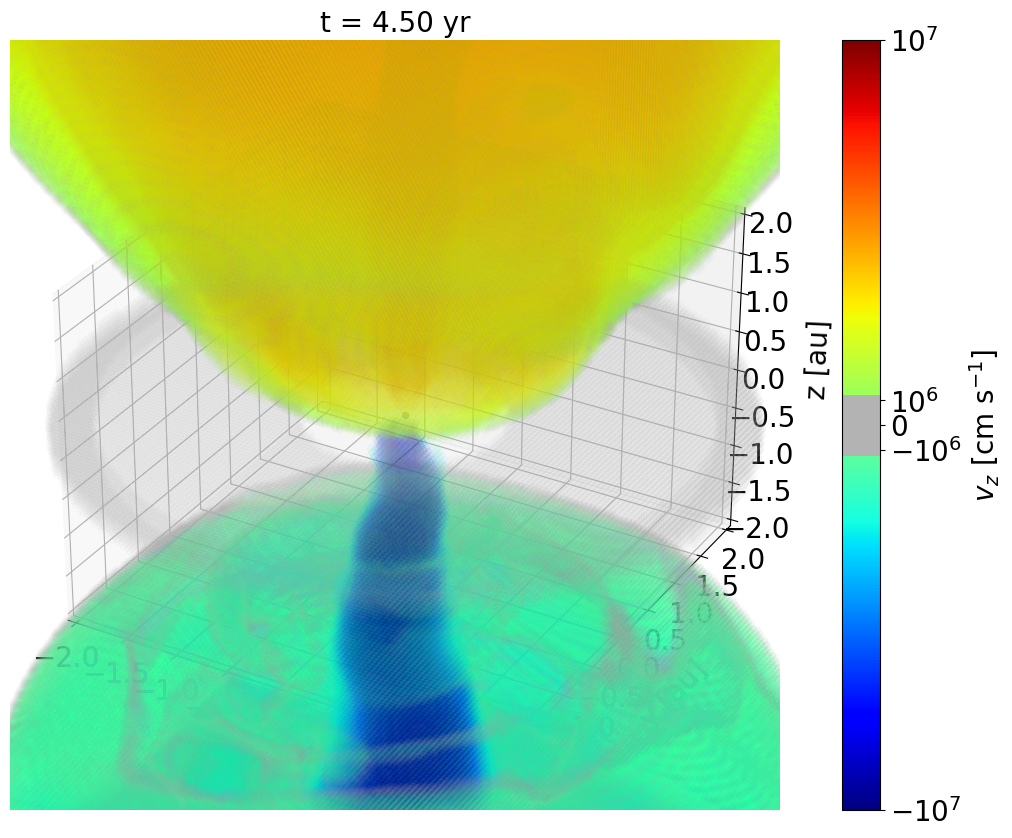}
    \caption{3D rendition of the jet, disk wind, and the disk at a representative time $t=4.5$~yr. The colors are the $\hat{z}$-direction velocity, showing a fast ($>100$~km/s) collimated bipolar jet (colored red and blue, close to the polar axis), surrounded by a slower disk wind (colored yellow and green, further away from the polar axis). The disk is represented by the gray structure on the equatorial plane. The jet is highly variable in space and time, as seen from the animated version available at \url{https://figshare.com/s/899801ea6282e9c06261?file=53753738}.
    The animated version is 50 seconds long, featuring a 3D rendition of the outflow in our model. We encourage readers to view it for a clearer impression of the outflow structure and dynamics.
    }
    \label{fig:jet3D}
\end{figure}

In this section, we present an overview of some broad features of the reference model \textsc{norot} with a non-rotating sink region; the \textsc{strot} model (with a rotating sink region) will be discussed in Appendix \ref{app:insensitivity}.  Fig.~\ref{fig:jet3D} displays a 3D view of the simulation at a representative time of $t=4.5$~yr, showing a bipolar outflow with a vertical velocity away from the disk larger than $10^6~\mathrm{cm\ s^{-1}}$. The disk is represented by the gray torus near the equatorial plane with density $>10^{-12}~\mathrm{g\ cm^{-3}}$. 
At this time, the narrow jet in the lower hemisphere is faster and more prominent than its counterpart in the upper hemisphere, although both jet components are highly variable in time, as can be seen from the animated version of the figure (see the figure caption for a link). Both jet components are surrounded by a wider, slower outflow -- the disk wind component. All models presented here cover a simulation time span of 10 years. While the system is highly variable and does not reach a true steady state, this duration is sufficient for the overall accretion and outflow properties to settle into a quasi-steady configuration and the key structural and dynamical features of the outflow remain qualitatively stable after $\sim$4~yr.

\begin{figure}
    \centering
    \includegraphics[width=\columnwidth]{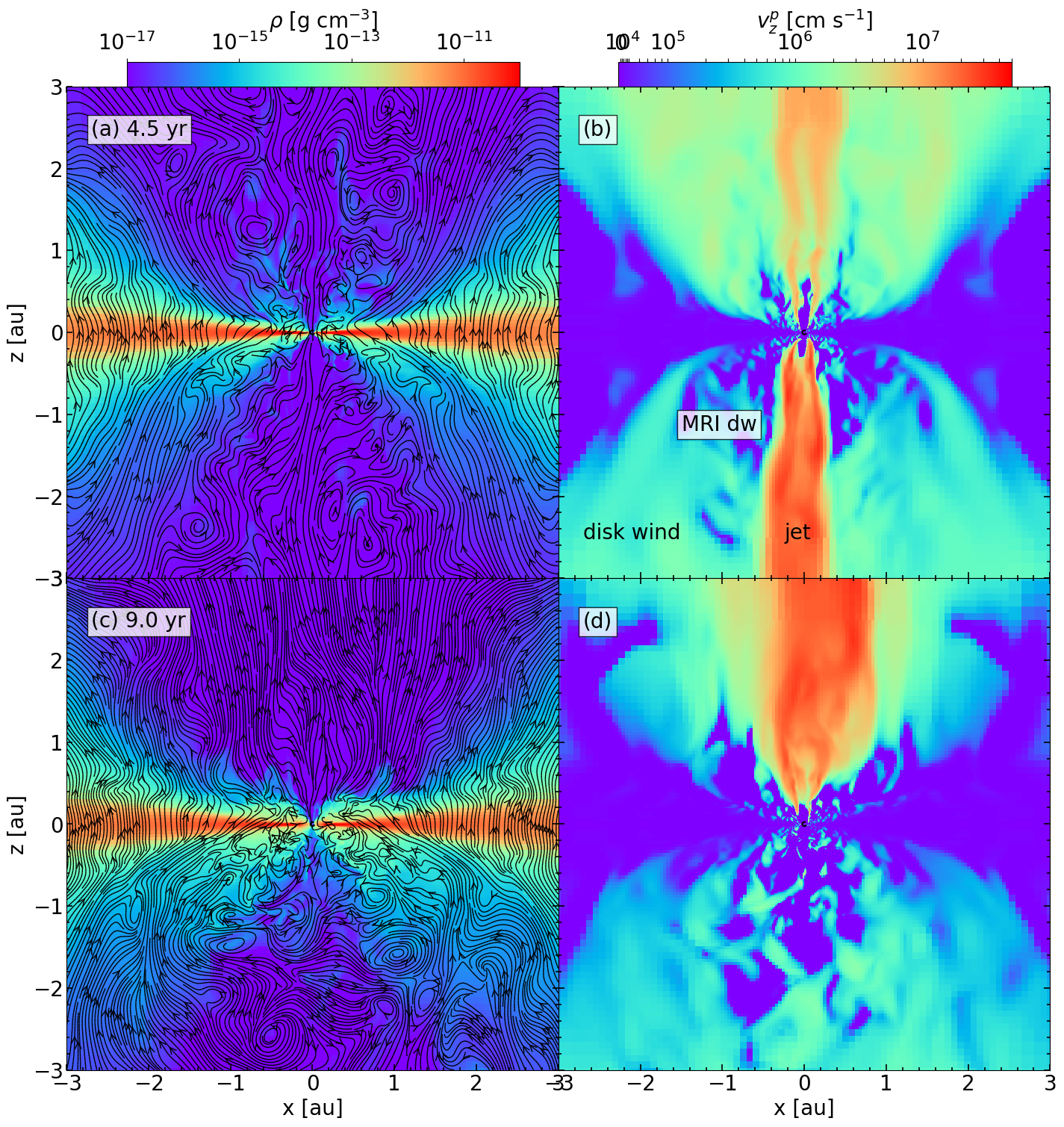}
    \caption{Overview of the simulation result, showing a slice through the plane at $y = 0$. The upper and lower panels show two representative times at 4.5 and 9.0~yr, respectively. Panels (a) and (c) show the density, and panels (b) and (d) show the projected $z$ direction velocity (equ.~\ref{equ:vz-projection}). The three main components of the outflow - jet, disk wind, and MRI-active disk wind (MRI DW) - are labeled in panel (b). An animated version is available at \url{https://figshare.com/s/8b13ca47b44adf6e3db9}.
    The animated version is 50 seconds long, showing an overview of our simulation result and the evolution of the outflow in our model.
    }
    \label{fig:overview}
\end{figure}

To show the outflow more quantitatively, Fig.~\ref{fig:overview} presents two $x$-$z$ slices at the two representative times: $t=4.5$ and $9.0$~yr. For each row, the horizontal panels show, respectively, the density $\rho$ and the projected z-direction velocity ($v_z^p$), defined as
\begin{equation}
    v_z^p = v_z~ z/|z|.
    \label{equ:vz-projection}
\end{equation}
A positive value of $v_z^p$ means that the flow is away from the disk midplane (i.e., an outflow).  The density is broadly symmetric about the midplane at both 4.5 and 9.0~yr, although there is significantly more mass in the polar region in the lower hemisphere at 9.0~yr. The fast jet, defined as the region where $v_z^p\geq 10^7~\mathrm{cm\ s^{-1}}$, is stronger in the lower hemisphere at 4.5~yr as mentioned earlier, but becomes stronger in the upper hemisphere at 9.0~yr. The reason behind this change in jet strength and its relationship to the density structure will be discussed in sec.~\ref{sec:jet_details}. Readers are encouraged to view the animated version of this figure to better visualize the evolution of our model, especially its strong time variability.

Surrounding the jet, but also moving away from the disk midplane, is a slower-moving disk wind. The disk wind is prominent at 4.5~yr as the wide-angle structure outflows on both hemispheres. The disk wind persists at 9.0~yr, particularly in the lower hemisphere, where a jet is absent. Between the jet and the disk wind lies an MRI-active disk wind region. We will focus on these three components one by one in the following subsections.

\subsection{The polar jet}
\label{sec:jet}

\begin{figure*}
    \centering
    \includegraphics[width=\linewidth]{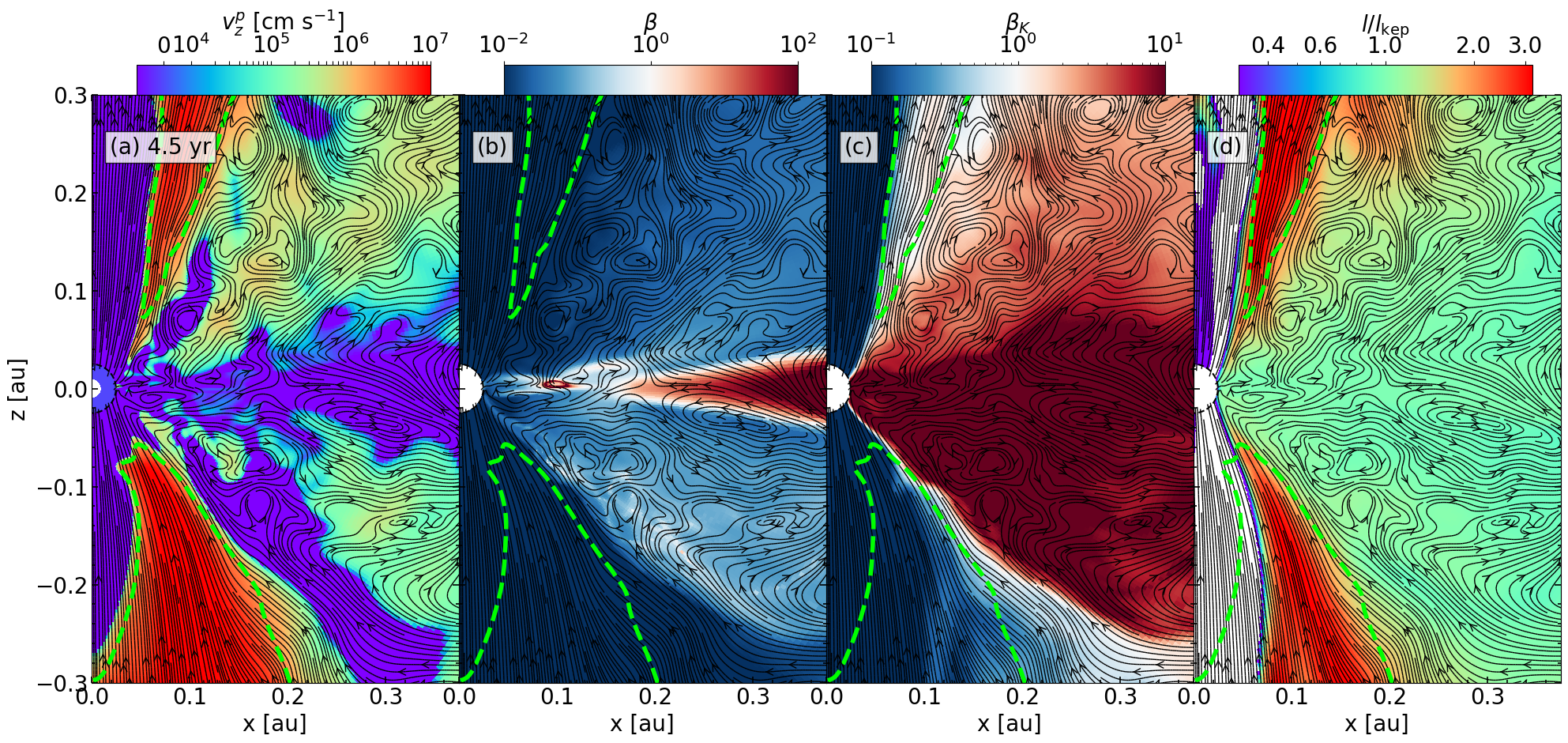}
    \caption{The gas properties near the base of the jet. Plotted are the azimuthally averaged projected $z$ direction velocity (equ.~\ref{equ:vz-projection}), plasma-$\beta$, kinetic-$\beta$ (equ.~\ref{equ:kinetic-beta}), and the comparison between the local specific angular momentum and the Keplerian value, respectively, at a representative time $t=4.5$~yr. The green contour highlights the lower part of the jet (where $v_z^p\ge 3\times10^6$~cm/s), where the plasma-$\beta$ is much less than unity, and the kinetic-$\beta$ is close to or less than unity. An animated version is available at \url{https://figshare.com/s/219f51de0c77cbcf17f1}.
    The animated version is 25 seconds long, showing an overview of the jet properties in our model.
    }
    \label{fig:r18_azimavg_props}
\end{figure*}
\begin{figure*}
    \centering
    \includegraphics[width=\linewidth]{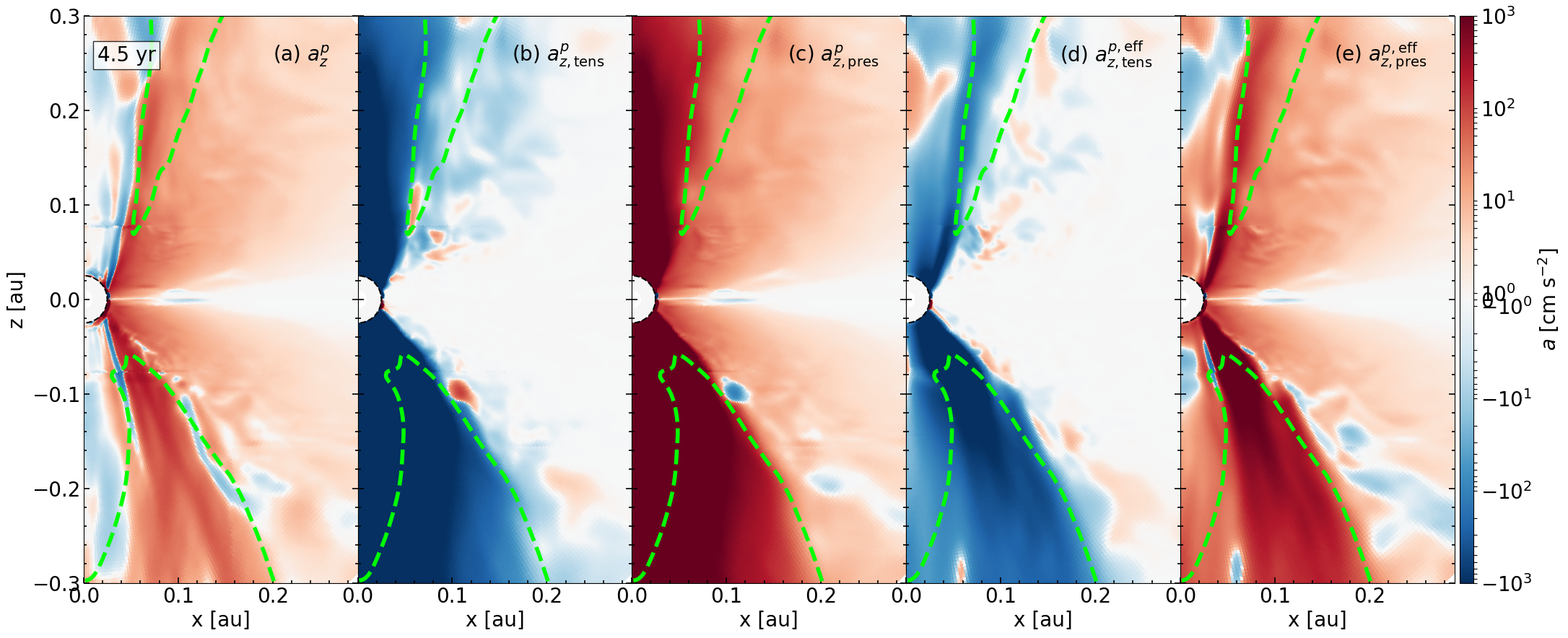}
    \caption{The gas acceleration near the base of the jet. The values are the $\hat{z}$-projected values, where a positive value means the force is pointing away from the disk midplane and a negative value points towards the midplane. The five panels show the total magnetic acceleration, gas acceleration due to magnetic tension, gas acceleration due to magnetic pressure, the gas acceleration due to the ``effective magnetic tension'' (defined in equ.~\ref{equ:az_mag_tens_eff}), and the gas acceleration due to the ``effective magnetic pressure'' (defined in equ.~\ref{equ:az_mag_pres_eff}). The outflow is launched by the toroidal magnetic pressure gradient, which provides the acceleration away from the disk in both the upper and lower hemispheres. Magnetic tension is unable to launch an outflow as it points towards the disk. An animated version is available at \url{https://figshare.com/s/d0d617e992b70dd720fb}.
    { The animated version is 50 seconds long, confirming that the magnetic forces remain consistent throughout the simulation and validating the representative frame shown here. It illustrates the dominant role of toroidal magnetic pressure in driving the outflow.}
    }
    \label{fig:r18_azimavg_forces}
\end{figure*}

As shown in panel (a) of Fig.~\ref{fig:overview}, the magnetic field lines threading the jet are connected to the inner active zone of the disk and the central boundary. Thus, we focus on these two regions to examine the jet-launching mechanism.

Fig~\ref{fig:r18_azimavg_props}(a) shows the azimuthally-averaged $v_{z}^p$ near the base of the jet-launching region. The green contour traces the jet in both the upper and lower hemispheres. The specific angular momentum in the jet launching region is well above the local Keplerian value as a result of angular momentum transport from the disk into the jet region. The importance of the magnetic field is evident: the plasma-$\beta$ is well below order unity (fig.~\ref{fig:r18_azimavg_props}[b]) in both the jet region and the magnetically active inner disk zone. Magnetic energy also dominates over kinetic energy in the jet region, as illustrated in  Fig~\ref{fig:r18_azimavg_props}(c), which shows that the kinetic-$\beta$
\begin{equation}
    \beta_K = \frac{\rho v^2/2}{B^2/8\pi},
    \label{equ:kinetic-beta}
\end{equation}
is below unity in the jet region, particularly in the lower hemisphere, where the outflow is stronger at 4.5~yr. In contrast, the kinetic-$\beta$ in the disk is above order unity, suggesting that the outflow-launching region contains two components: a kinetic energy-dominated disk and a magnetically dominated jet-launching region. 

To understand the jet production mechanism in the magnetically-dominated jet-launching region, we show in Fig.~\ref{fig:r18_azimavg_forces} the $\hat{z}$-direction projected total magnetic acceleration ($a_z^p$, panel [a]), magnetic tension acceleration ($a_{z, \mathrm{tens}}^p$, panel [b]), magnetic pressure acceleration ($a_{z, \mathrm{pres}}^p$, panel [c]), and the projected $\hat{z}$-direction magnetic acceleration due to poloidal magnetic tension (panel [d]) and toroidal magnetic field gradient (panel [e]). 
The projected $z$-direction magnetic acceleration in cylindrical coordinates is defined as
\begin{equation}
    a_z^p = \frac{z}{|z|}\Big[\frac{1}{4\pi\rho}(\nabla\times\mathbfit{B})\times\mathbfit{B}\Big]_z;
    \label{equ:az_mag_force}
\end{equation}
\begin{equation}
    \begin{split}
        a_{z, \mathrm{tens}}^p &= \frac{z}{|z|}\Big[\frac{1}{4\pi\rho}(\nabla\cdot\mathbfit{B})\mathbfit{B}\Big]_z \\
        &= \frac{z}{|z|}\frac{1}{4\pi\rho}
        \Big[B_R\frac{\partial B_z}{\partial R} + B_\phi\frac{\partial B_z}{\partial \phi} + B_z\frac{\partial B_z}{\partial z}\Big];
    \end{split}
    \label{equ:az_mag_tens}
\end{equation}
\begin{equation}
    \begin{split}
    a_{z, \mathrm{pres}}^p &= -\frac{z}{|z|}\Big[\frac{1}{8\pi\rho}(\nabla\cdot\mathbfit{B}^2)\Big]_z \\
    &= -\frac{z}{|z|}\frac{1}{4\pi\rho}\Big[B_R\frac{\partial B_R}{\partial z} + B_\phi\frac{\partial B_\phi}{\partial z} + B_z\frac{\partial B_z}{\partial z}\Big],
    \end{split}
    \label{equ:az_mag_pres}
\end{equation}
where the $\frac{z}{|z|}$ is the projection in the direction away from the disk midplane. Note that the $z$-direction magnetic acceleration term $B_z\frac{\partial B_z}{\partial z}$ is in both the magnetic tension acceleration (equ.~\ref{equ:az_mag_tens}) and the magnetic pressure acceleration (equ.~\ref{equ:az_mag_pres}) equations, and has opposite signs in these two equations, which means the {\it net} effect of the $B_z\frac{\partial B_z}{\partial z}$ term is zero in launching of the outflow. Thus we can rewrite equ.~\ref{equ:az_mag_pres} excluding this term to define an ``effective magnetic pressure acceleration''
\begin{equation}
    \begin{split}
        a_{z, \mathrm{pres}}^{p, \mathrm{eff}} &= -\frac{z}{|z|}\frac{1}{4\pi\rho}\Big[B_R\frac{\partial B_R}{\partial z} + B_\phi\frac{\partial B_\phi}{\partial z}\Big] \\
        &\approx -\frac{z}{|z|}\frac{1}{4\pi\rho} \frac{\partial B_\phi}{\partial z}B_\phi,
    \end{split}
    \label{equ:az_mag_pres_eff}
\end{equation}
The last approximation uses the fact that the toroidal field dominates the outflow region. A similar argument can be applied to equ.~\ref{equ:az_mag_tens} to define an ``effective magnetic tension acceleration''
\begin{equation}
    a_\mathrm{z, tens}^{p, \mathrm{eff}} = \frac{z}{|z|}\frac{1}{4\pi\rho}B_R \frac{\partial B_z}{\partial R},
    \label{equ:az_mag_tens_eff}
\end{equation}
assuming azimuthal symmetry. 

Fig.~\ref{fig:r18_azimavg_forces}(d) and (e) illustrate these two effective accelerations. While $a_{z, \mathrm{tens}}^{p, \mathrm{eff}}$ always acts toward the disk and therefore cannot drive the jet, $a_{z, \mathrm{pres}}^{p, \mathrm{eff}}$ is directed away from the disk, playing a crucial role in jet acceleration. It overcomes $a_{z, \mathrm{tens}}^{p, \mathrm{eff}}$ and contributes positively to the net magnetic acceleration ($a_z$), as shown in Fig.~\ref{fig:r18_azimavg_forces}(a). A geometric argument for why only magnetic pressure, not magnetic tension, can accelerate a jet will be presented in sec.~\ref{fig:r18_azimavg_forces}. We stress here that the jet is primarily accelerated by the toroidal magnetic pressure gradient.

\subsection{The disk wind}
\label{sec:diskwind}
Surrounding the jet but further away from the polar axis is a slower disk wind (e.g. fig.~\ref{fig:overview}[b]). The $z$-direction gas velocity in the disk wind is around $10^6\mathrm{cm\ s^{-1}}$, which is slower than the jet. In this section, we focus on the launching mechanism of the disk wind.

\begin{figure*}
    \centering
    \includegraphics[width=\linewidth]{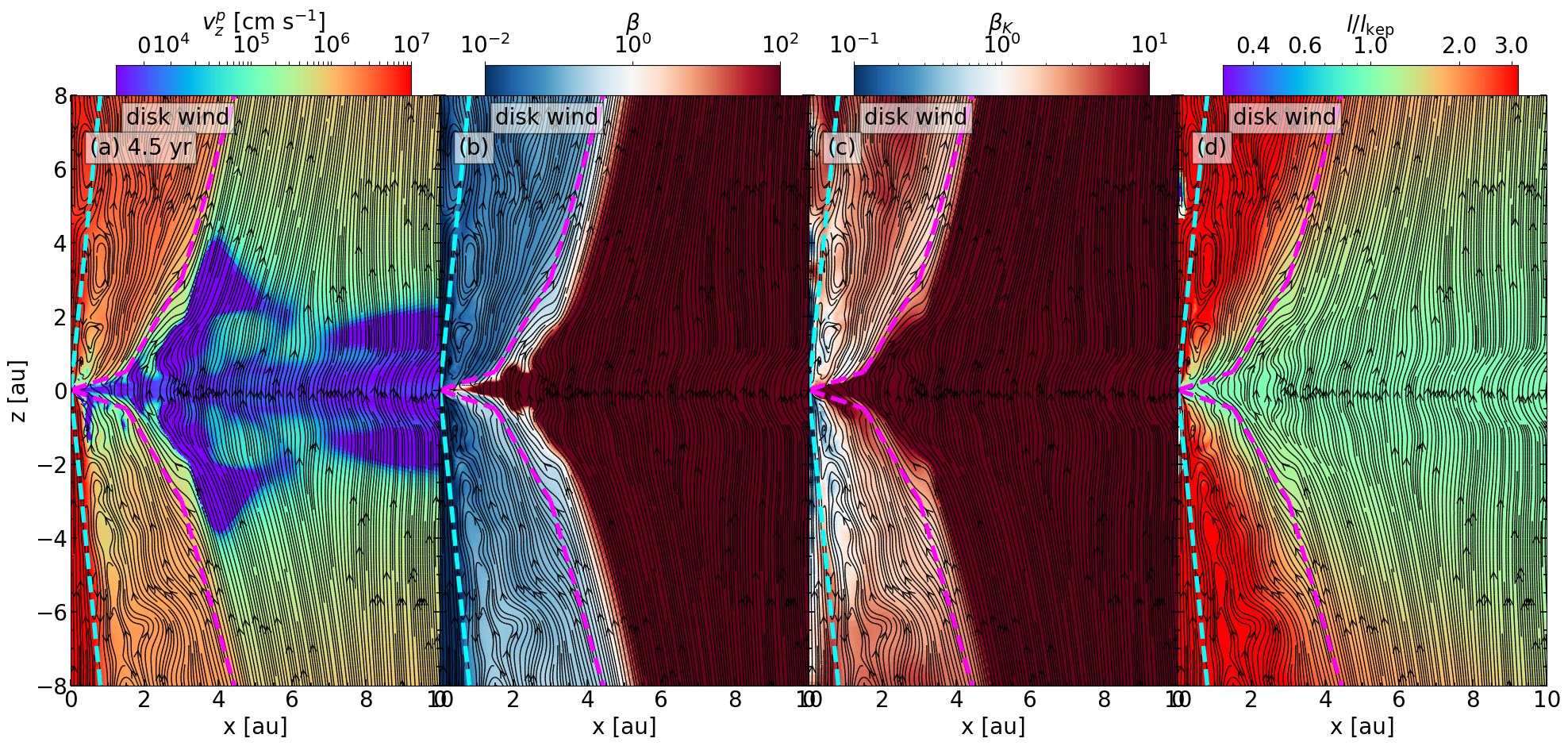}
    \caption{The gas properties of the disk wind. The four panels show the same quantities as in fig.~\ref{fig:r18_azimavg_props}, at the same representative time but on a larger scale. The cyan line shows the approximate boundary between the disk wind and the jet, and the magenta line shows the approximate boundary between the disk wind and the expanded disk atmosphere. Compared to the jet, the disk is less magnetically dominated as the kinetic-$\beta$ is above order unity. An animated version is available at 
    \url{https://figshare.com/s/668556d9b7090225b367}.
    { The animated version is 25 seconds long, showing an overview of the disk wind launched in our model.}
    }
    \label{fig:r18_azimavg_outer_vz_betas_l}
\end{figure*}

Fig.~\ref{fig:r18_azimavg_outer_vz_betas_l}(a) shows the $v_z^p$ in the disk wind region. 
The disk wind is the region between the cyan dashed line (the approximate boundary between the jet and the disk wind) and the magenta line (the boundary between the disk wind and the expanded disk atmosphere).
Fig.~\ref{fig:r18_azimavg_outer_vz_betas_l}(b) and (c) show the plasma-$\beta$ and kinetic-$\beta$ in the disk wind region respectively. While plasma-$\beta$ is still well below order unity in the disk wind region, the kinetic-$\beta$ is above unity. Unlike the jet, which is completely magnetically dominated, the kinetic energy (primarily from rotation) is important in the disk wind region. Similar to the jet, the specific angular momentum in the disk wind region exceeds the local Keplerian value (by a factor of at least 3, see Fig.~\ref{fig:r18_azimavg_outer_vz_betas_l}[d]).

\begin{figure*}
    \centering
    \includegraphics[width=\linewidth]{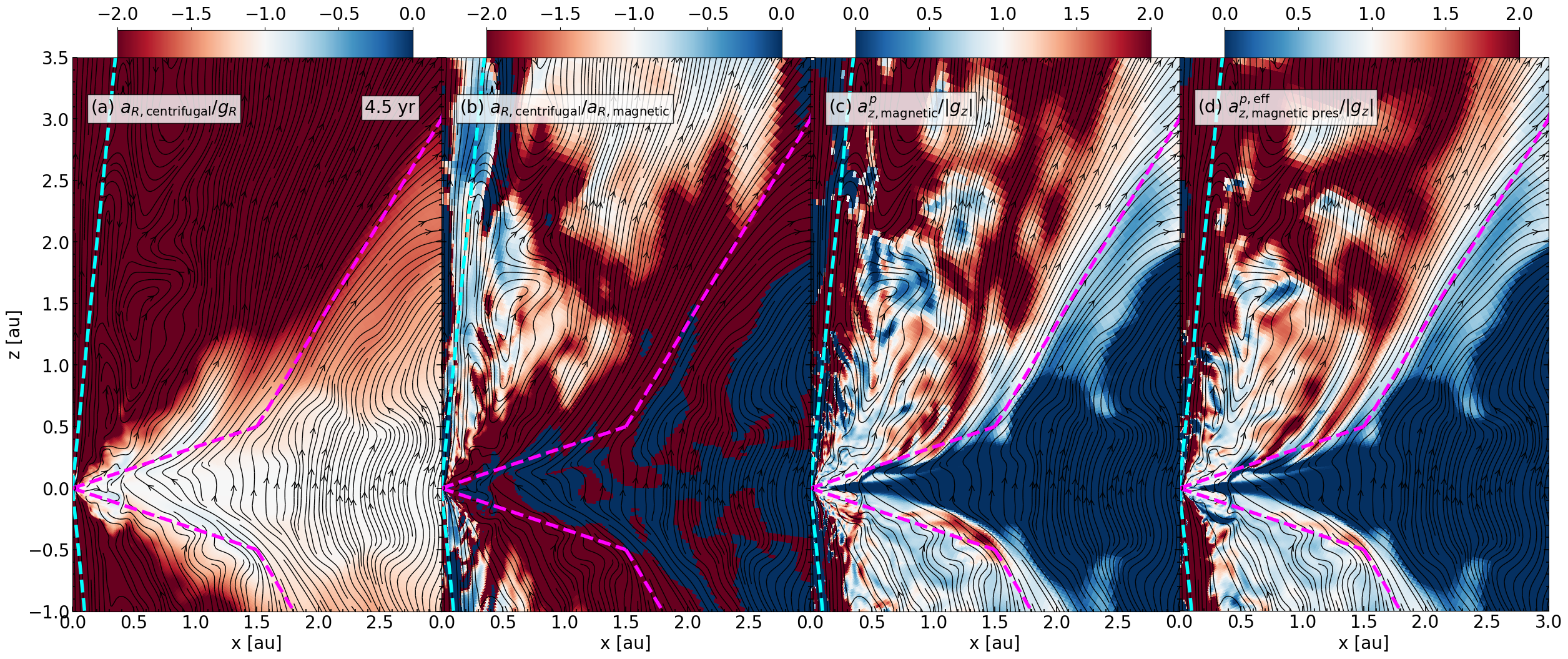}
    \caption{A comparison between different forces in the disk wind. \textbf{Panel (a)} shows the ratio between the centrifugal acceleration and the gravitational acceleration in the cylindrical-$\hat{R}$ direction; \textbf{panel (b)} shows the ratio between the centrifugal acceleration and magnetic acceleration. Negative values in panels (a) and (b) indicate that the centrifugal acceleration points in the opposite direction to both gravity and the magnetic field. \textbf{Panel (c) and (d)} show the $\hat{z}$ direction magnetic acceleration and $\hat{z}$ direction ``effective'' magnetic pressure acceleration (defined in equ.~\ref{equ:az_mag_pres_eff}) normalized by gravitational acceleration in the $\hat{z}$ direction respectively. Toroidal magnetic pressure drives the disk wind upwards in the $\hat{z}$ direction. An animated version is available at 
    \url{https://figshare.com/s/86c130e6b25e55d81200}.
    The animated version is 50 seconds long, showing the force ratios in the disk wind are similar to those in the representative frame we show here throughout the simulation.
    }
    \label{fig:r18_azimavg_outer_forces}
\end{figure*}

Because both magnetic and kinetic energy are important for the disk wind, we will examine both the acceleration by the magnetic field and by momentum (centrifugal force) to understand the acceleration of the disk wind. Fig.~\ref{fig:r18_azimavg_outer_forces} shows the centrifugal force in the cylindrical-$\hat{R}$ direction ($a_{R, \mathrm{centrifugal}}$), magnetic force in the cylindrical-$\hat{R}$ direction ($a_{R, \mathrm{magnetic}}$) and $\hat{z}$ directions, and effective magnetic pressure force (equ.~\ref{equ:az_mag_pres_eff}) respectively. In the disk wind region, $a_{R, \mathrm{centrifugal}}$ opposes both the gravitational acceleration ($g_R$) and $a_{R, \mathrm{magnetic}}$, which direct inward. The magnitude of $a_{R, \mathrm{centrifugal}}$ exceeds the $g_R$ by more than a factor of 2 and generally dominates over $a_{R, \mathrm{magnetic}}$. This outward dominance of centrifugal force over both the magnetic field and gravity (as shown in fig.~\ref{fig:r18_azimavg_outer_vz_betas_l}[d], where the gas is super-Keperian in the disk wind region) drives the gas horizontally outward. In the $\hat{z}$ direction, magnetic force provides a $\hat{z}$-direction acceleration away from the disk midplane against gravity (fig.~\ref{fig:r18_azimavg_outer_forces}[c]). The net $\hat{z}$ direction acceleration is comparable to the net cylindrical-$\hat{R}$ direction acceleration, so the net acceleration moves diagonally on a meridional plane, which produces the wide-angle disk wind.

Similar to the jet, toroidal magnetic pressure is the main driver of the disk wind. Fig.~\ref{fig:r18_azimavg_outer_forces}(d) shows the effective $\hat{z}$-direction magnetic acceleration (equ.~\ref{equ:az_mag_pres_eff}), which is very similar to the total $\hat{z}$-direction magnetic acceleration (Fig.~\ref{fig:r18_azimavg_outer_forces}[c]). A similar geometric argument can be used to illustrate why only magnetic pressure can drive a $z$-direction acceleration, and we leave the detailed illustration to sec.~\ref{sec:geometry}. 

\subsection{The MRI-active turbulent inner disk wind}

\label{sec:middle zone}
\begin{figure}
    \centering
    \includegraphics[width=0.7\linewidth]{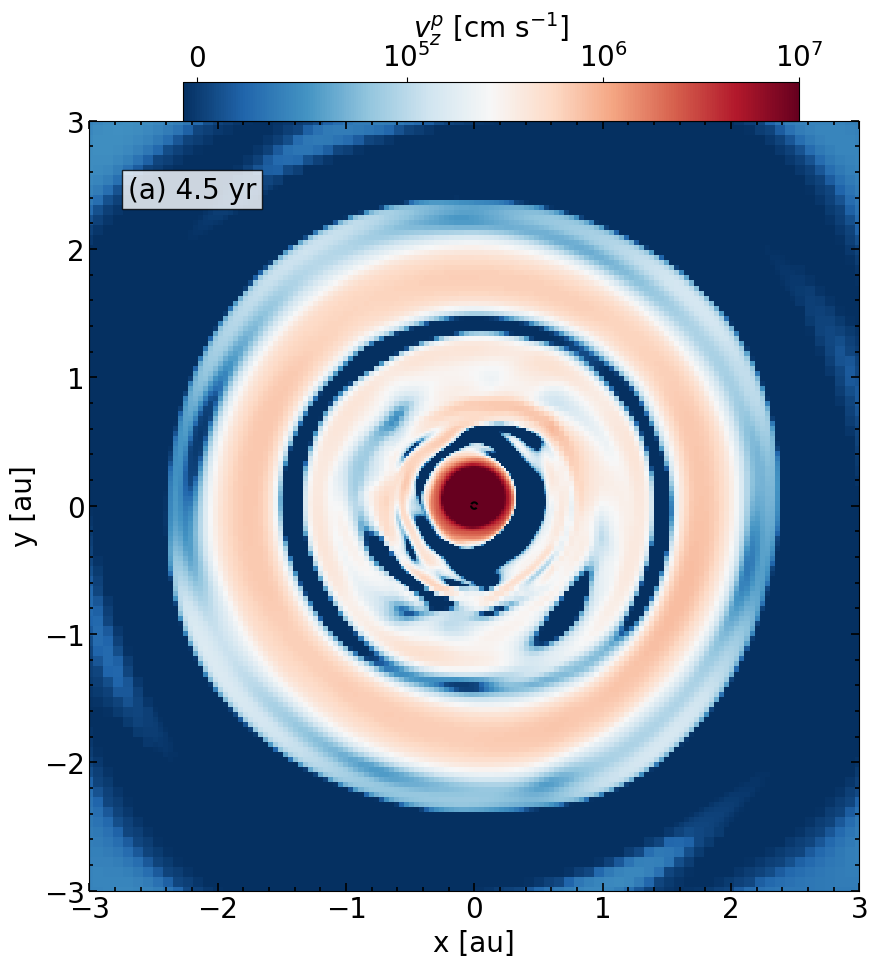}
    \caption{A $x$-$y$ slice through the simulation domain at $z = -1.0$~au showing the projected $z$-direction velocity (equ.~\ref{equ:vz-projection}). The jet stands out at the center of the slice as the dark red (with $v_z^p \ge 10^7$~cm/s) circle with radius $\sim 0.3$~au; the disk wind is the annulus in lighter red at $\sim 1.5$~au; in between the jet and the disk wind is the MRI-active turbulent disk wind, where the gas alternates between infall toward the disk and outflow. An animated version is available at \url{https://figshare.com/s/5d5a3c8cb83a94170519}.
    The animated version is 50 seconds long, showing the outflow structure described above is persistent in our model before the MRI-active turbulent disk wind chokes the jet at later times.}
    \label{fig:r18_upperzslice_vzp}
\end{figure}

The inner portion of the disk wind at low altitudes (e.g., near $(x, z) = (1, 1)$ au in fig.~\ref{fig:r18_azimavg_outer_forces}) is a region where the magnetic force alternates between dominating gravity in the vertical direction and being less than it. To show the difference between this region and the jet and the disk wind more vividly, Fig.~\ref{fig:r18_upperzslice_vzp} shows a horizontal slice of the $v_z^p$ at $z = -1.0$~au. The jet stands out at the center as the circular structure moving at a speed $>10^7\ \mathrm{cm\ s^{-1}}$. The disk wind is the slower outward-moving ring located at $R \approx 1.5\sim 2~\mathrm{au}$. Between the jet and the disk wind is a region with slow outflow {\it and} slow infall. This middle region, an MRI-active turbulent disk wind region, separates the jet from the outer laminar disk wind.

\begin{figure*}
    \centering
    \includegraphics[width=\linewidth]{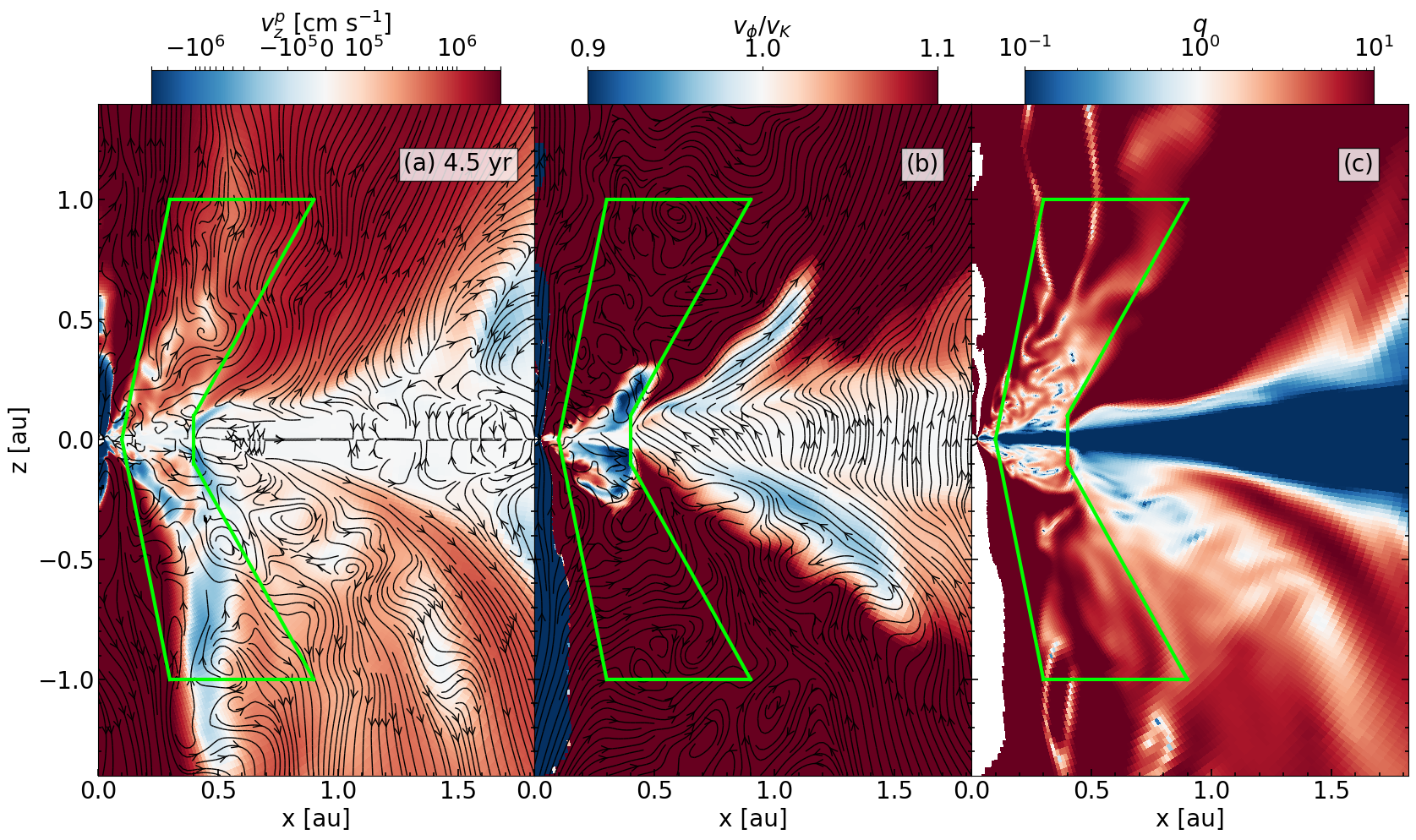}
    \caption{Properties of the MRI-active turbulent disk wind region, enclosed by the green line. \textbf{Panel (a)} shows the projected $z$-direction velocity (equ.~\ref{equ:vz-projection}), overplotted with velocity streamlines; \textbf{panel (b)} shows the ratio between local rotation velocity $v_\phi$ and the local Keplerian velocity $v_K$, overplotted with magnetic field lines. \textbf{Panel (c)} shows the normalized most unstable MRI wavelength (equ.~\ref{equ:mri_wavelength}), where $10^{-1} < q < 10^1$ is the most optimal range to activate MRI. An animated version is available at \url{https://figshare.com/s/2a211268b205eb9a2dd1}.
    The animated version is 25 seconds long, showing the evolution of the MRI active turbulent disk wind region in our model. The highlighted features in each panel are persistent in the animation.
    }
    \label{fig:r18_middle_region}
\end{figure*}

To illustrate the physical origin of the MRI-active turbulent disk wind region, we show the azimuthally-averaged $v_z^p$ in Fig.~\ref{fig:r18_middle_region}(a). The green lines denote rough boundaries between the jet, the middle MRI-active turbulent disk wind region, and the more laminar disk wind at a larger cylindrical radius. The velocity streamlines are overplotted in Fig.~\ref{fig:r18_middle_region}, showing that the middle region identified in Fig.~\ref{fig:r18_upperzslice_vzp} originates around the boundary between the dead zone and active zone of the disk. Fig.~\ref{fig:r18_middle_region}(b) shows the ratio between the $\hat{\phi}$-direction velocity and the local Keplerian velocity, where the base of the middle region stands out as the sub-Keplerian region sandwiched between the super-Keplerian jet and super-Keplerian disk wind. Because the velocity streamlines are more chaotic in the middle region than in both the jet and the disk wind, the magnetic field lines (shown in Fig.~\ref{fig:r18_middle_region}[b]) are also more tangled in the MRI-active turbulent disk wind region. This disordered magnetic fieldline means that the usual expectation that angular momentum can be transferred along an (ordered) magnetic field line to gas at a larger cylindrical radius cannot be met in the MRI-active turbulent disk wind region. Thus, the disk surface in the MRI-active turbulent disk wind region expands outward as a result of magnetic pressure without gaining angular momentum, resulting in the sub-Keplerian region.

The reason for the chaotic gas flow pattern and the disordered fieldline in the middle region is that this region is unstable to turbulence induced by magneto-rotational instability (MRI). The most unstable turbulent MRI wavelength is given by \citep[][]{Noble2010}
\begin{equation}
    \lambda=2\pi\sqrt{\frac{16}{15}}\frac{v_A}{\Omega}.
\end{equation}
For MRI to be active, the wavelength must not be too long such that the MRI wavelength cannot fit into the disk \citep{Porth2021}, and the wavelength must also not be too short such that the MRI effect is too weak \citep{Balbus1991, Hawley1991}. To identify the region where MRI is active, we first define a dimensionless wavelength 
\begin{equation}
    q = \frac{\lambda}{H} \approx \frac{\lambda}{R/10},
    \label{equ:mri_wavelength}
\end{equation}
where $H$ is the scale height of the disk, which can be approximated by one-tenth of the cylindrical radius $R$. The MRI operates most efficiently when $q$ is of order unity, which motivates us to adopt, for illustrative purposes, the range $0.1 < q < 10$ where the most unstable wavelength is neither too long nor too short. 

Fig.~\ref{fig:r18_middle_region}(c) shows $q$: while the jet, the disk wind, and the disk are all MRI stable ($q > 10$ in the disk wind and the jet; $q < 0.1$ in the disk), the MRI-active turbulent disk wind region lies within the optimal range of MRI. Because this inner disk wind region is MRI-active, both velocity and magnetic field are far from laminar. Thus, although the gas is still moving away from the disk due to the magnetic pressure gradient (Fig.~\ref{fig:r18_azimavg_forces}[c]), an ordered outflow cannot be launched in this region. 

The lack of an ordered outflow at a low altitude does not translate to the complete absence of an outflow. When MRI cannot be sustained at a higher altitude, the MRI-active turbulent disk wind becomes a laminar, ordered disk wind. This transition is illustrated in Fig.~\ref{fig:r18_middle_region}(a), where the gas above the region enclosed by the green line is more laminar than the gas within due to MRI not being able to be sustained at a higher altitude (Fig.~\ref{fig:r18_middle_region}[c]). The gas, after the transition from turbulent to laminar, becomes part of the slower disk wind surrounding the jet.

In summary, both the jet and the disk wind are driven by toroidal magnetic pressure. The region responsible for launching the jet is primarily governed by the magnetic field, while the disk wind-launching region is dominated by kinetic energy, with centrifugal forces acting to widen the opening angle of the disk wind. The jet and disk wind are distinct outflows, separated by an MRI-active turbulent disk wind zone, which becomes part of the disk wind at a large altitude when the MRI-induced turbulence can no longer be sustained.

\section{A lightly-loaded toroidal magnetic field pressure-driven jet}
\label{sec:jet_details}

\begin{figure}
    \centering
    \includegraphics[width=\linewidth]{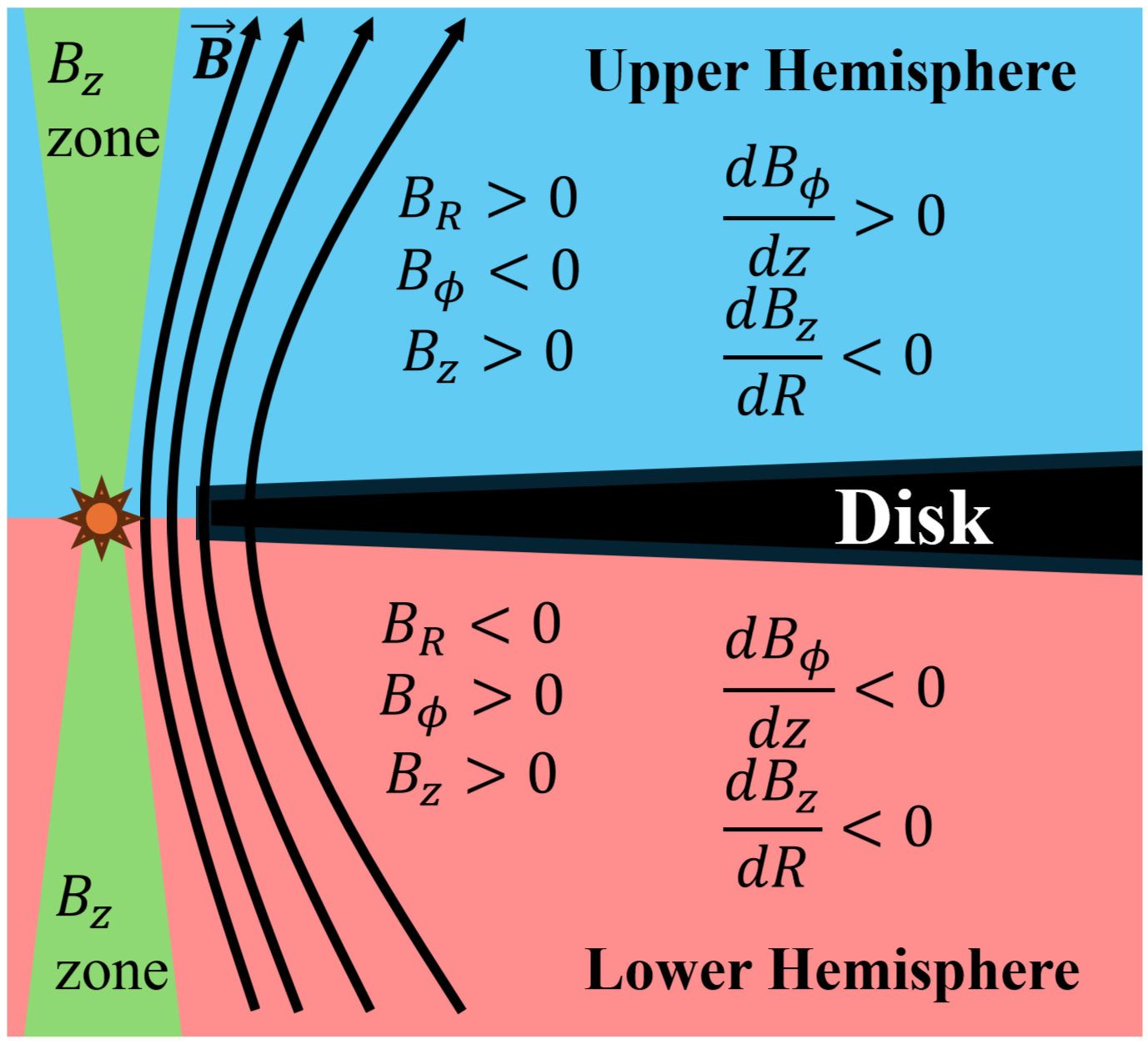}
    \caption{Geometric argument for the magnetic field geometry (sec.~\ref{sec:geometry}), assuming the disk is threaded by open, $+\hat{z}$-direction magnetic field lines. The signs of the magnetic field and two magnetic field gradients above and below the disk in cylindrical coordinates are shown for the upper and lower hemispheres, respectively, showing magnetic tension force (equ.~\ref{equ:az_mag_tens_eff}) is unable to drive an outflow, whereas magnetic pressure force (equ.~\ref{equ:az_mag_pres_eff}) can.}
    \label{fig:geometry}
\end{figure}

\subsection{A geometric argument for magnetic field structure above and below the disk}
\label{sec:geometry}

In this section, we present a geometric argument that an outflow can only be launched by a toroidal field pressure gradient, and tension cannot launch an outflow in a global model.

The geometry of the magnetic field is illustrated in Fig.~\ref{fig:geometry} in the outflow-launching region. Assuming global $B_z > 0$, $B_\phi < 0$ in the upper hemisphere because the disk has a faster speed compared to the atmosphere; $\frac{dB_\phi}{dz} > 0$, the strength of (the negative) toroidal magnetic field decreases with height due to the decrease of differential rotation at a higher altitude. Thus, the effective magnetic pressure force (equ.~\ref{equ:az_mag_pres}), $a_\mathrm{z, pres}^{p, \mathrm{eff}} > 0$, capable of launching an outflow.

The effective magnetic tension force (equ.~\ref{equ:az_mag_tens_eff}), however, always points towards the disk midplane. $B_R > 0$ in the upper hemisphere both because of geometry in the jet-launching region (Fig.~\ref{fig:r18_azimavg_props}) and the excess of specific angular momentum in the disk wind region (Fig.~\ref{fig:r18_azimavg_outer_vz_betas_l}[d]); $\frac{dB_z}{dR} < 0$, the strength of $B_z$ decreases as cylindrical radius increases. As a result, $a_\mathrm{z, tens}^{p, \mathrm{eff}} < 0$, which points to the midplane and is unable to launch an outflow. 

This geometric argument and the force analysis (Fig.~\ref{fig:r18_azimavg_forces}) both show that toroidal magnetic pressure launches an outflow (the jet and the disk wind). The only role of rotation is to generate the $B_\phi$ in the mass-dominated disk required to launch the outflow. Using the same geometric argument, we can show that $B_\phi < 0$ in the upper hemisphere and $B_\phi > 0$ in the lower hemisphere. The $B_\phi$ in the disk varies in the simulation, determining the location of the ``avalanche accretion stream''--a crucial component in jet launching. 

\subsection{$B_\phi$ in the disk}

\begin{figure*}
    \centering
    \includegraphics[width=\textwidth]{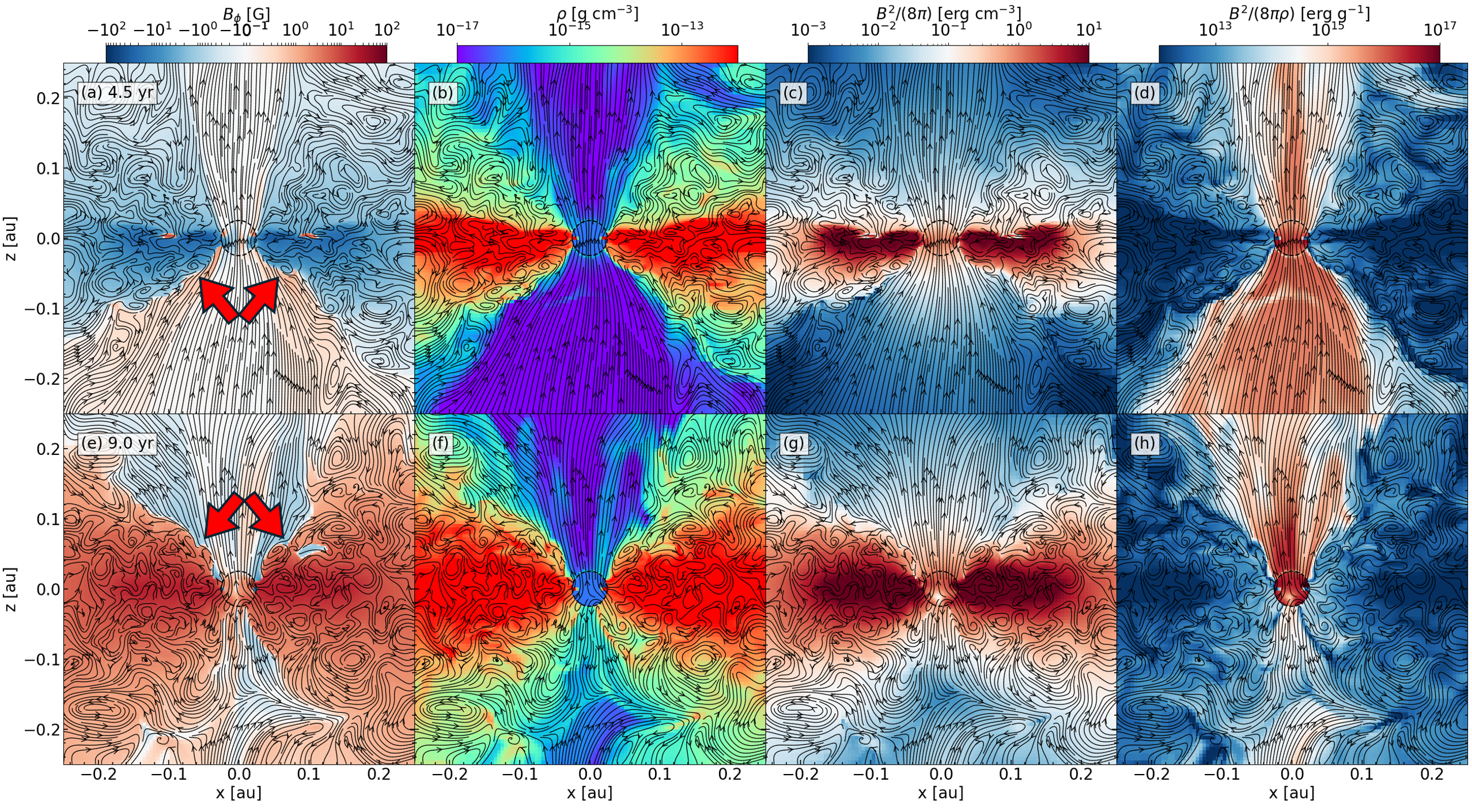}
    \caption{Properties of the disk surface avalanche accretion stream and the jet-launching cavity at two representative times: 4.5 yr (upper panels) and 9.0 yr (lower panels). The arrows in the two leftmost panels point to the locations of the avalanche accretion stream. \textbf{Panels (a) and (e)} show $B_\phi$, which switches direction inside the disk between 4.5 yr and 9.0 yr. \textbf{Panels (b) and (f)} display the gas density distribution. The disk surface avalanche accretion stream is located where $B_\phi$ switches sign spatially, roughly coinciding with the pinching of poloidal magnetic field lines. Low-density outflow cavities exist in both hemispheres at 4.5 yr (larger in the lower hemisphere), but much more prominently in the upper hemisphere at 9.0 yr, where $B_\phi$ spatially changes sign. \textbf{Panels (c) and (g)} illustrate the magnetic energy density, symmetric about the disk midplane at both times. \textbf{Panels (d) and (h)} show the magnetic energy per unit mass, which is large only in the low-density outflow cavities.
    An animated version can be found at \url{https://figshare.com/s/763d65cae5aa7d452442}.
    The animated version is 50 seconds long, showing the evolution of the magnetic field structure and magnetic energy density evolution in our model. The animation highlights the role of the avalanche accretion stream, located where $B_\phi$ switches sign, in limiting the disk expansion and facilitating jet launching.
    } 
    \label{fig:avalanche_accretion}
\end{figure*}

We have established in sec.~\ref{sec:middle zone} that the inner disk region is turbulent due to MRI. One consequence of being turbulent is that $B_\phi$ appears to take only one sign in most of the inner disk, probably because rapid magnetic reconnection due to MRI turbulence tends to remove $B_\phi$ of the opposite direction quickly, particularly in the inner disk region where plasma-$\beta < 1$ (fig.~\ref{fig:r18_azimavg_props}). This is illustrated in Fig.~\ref{fig:avalanche_accretion} panels (a) and (e), which show $B_\phi$ at 4.5 and 9.0 yr, respectively. At 4.5 years, $B_\phi$ is almost universally negative in the disk (albeit a negligible positive $B_\phi$ region exists), and at 9.0 years, $B_\phi$ switches to fully positive in the disk. 

The sign change in the disk from 4.5 to 9.0~yr is due to one exception to this one-sign rule: the disk $B_\phi$ can switch sign when $B_\phi$ changes sign at the region where $B_\phi$ is most rapidly generated, located at the innermost disk around the inner boundary. In this case, the $B_\phi$ in the disk would gradually switch to the opposite sign as the $B_\phi$ of the opposite sign slowly takes over the entire disk. Readers are encouraged to view the animated version of Fig.~\ref{fig:avalanche_accretion} for the transition from negative $B_\phi$ in the disk at 4.5 yr to positive $B_\phi$ at 9.0 yr.

\subsection{Avalanche accretion streams}
\label{sec:avalanche accretion}

The avalanche accretion stream is most prominent where the azimuthal magnetic field, $B_\phi$, changes sign. Figure~\ref{fig:avalanche_accretion} illustrates such streams at 4.5 and 9.0~yr in the upper and lower rows, respectively. The dense avalanche accretion stream resides where $B_\phi$ spatially changes sign and the meridian magnetic field line sharply pinches (highlighted by the arrows in the two leftmost panels). Because the avalanche accretion stream is a consequence of magnetic braking that removes angular momentum \citep[][]{Zhu2018, Tu2024c}, the locations where the azimuthal magnetic field pinches (where $B_\phi$ changes sign) are the most ideal for forming an avalanche accretion stream. Magnetic energy is generated in the disk by differential rotation, producing the toroidal magnetic field, with a magnetic energy density similar in both hemispheres of the disk atmosphere at 4.5 and 9.0~yr (Fig.~\ref{fig:avalanche_accretion}[c], [g]). However, while the overall magnetic energy density remains comparable, the magnetic energy per unit mass is more abundant in the upper hemisphere at 9.0~yr, resulting in the differences in the outflows launched from each hemisphere.

The MRI-active inner disk can expand vertically because of the toroidal magnetic pressure. As illustrated in Fig.~\ref{fig:r18_middle_region}, this expansion can reach a high altitude. In some extreme cases, this expansion can fill the entire disk atmosphere, including the jet-launching zone around the polar region, which can choke the jet by loading too much mass onto the magnetic fields. This is the case in the lower hemisphere at 9.0~yr, where the polar region is filled with high-density gas (Fig.~\ref{fig:avalanche_accretion}[f]), and no fast jet is launched (Fig.~\ref{fig:overview}). In the upper hemisphere at 9.0 years, however, the avalanche accretion stream stops the disk expansion. Because $B_\phi$ must change sign somewhere in the upper hemisphere, the magnetic field line is pinched azimuthally where $B_\phi = 0$, resulting in strong magnetic braking and facilitating a fast infall. This infall stream naturally separates the heavily mass-loaded raised inner disk atmosphere (where kinetic-$\beta > 1$) and the lightly loaded jet-launching zone in the polar region (where kinetic-$\beta < 1$) near the rotation axis through the formation of an avalanche accretion stream that typically starts somewhere above the inner active zone and ends close to the central radial boundary. The separation reduces the mass-loading in the jet-launching region, allowing a strong magnetic field to accelerate a small amount of gas mass there (Fig.~\ref{fig:avalanche_accretion}[h]), forming a jet.

The absence of $B_\phi$ switching sign in the disk atmosphere, which is necessary for forming an avalanche-accretion stream, is less favorable for jet launching. However, a jet can still form through a similar magnetic-pressure driven mechanism if the polar region remains lightly loaded—that is, if it is not filled with mass lifted by the inner disk MRI-induced turbulence. This is the case for the upper hemisphere at 4.5~yr (Fig.~\ref{fig:avalanche_accretion} upper row): although $B_\phi$ is ubiquitously negative there, the polar region is not filled with dense gas, so a jet can still be launched. It is, however, less developed (i.e. slower) than the jet in the lower hemisphere at the same time (see, e.g., Fig.~\ref{fig:overview}[b]), where a prominent avalanche stream with $B_\phi$ reversal separates the dense disk atmosphere and much less dense jet zone. 

\subsection{Jet from magnetic energy release in the polar cavity}
\label{sec:jet_as_energy_release}

\begin{figure*}
    \centering
    \includegraphics[width=\linewidth]{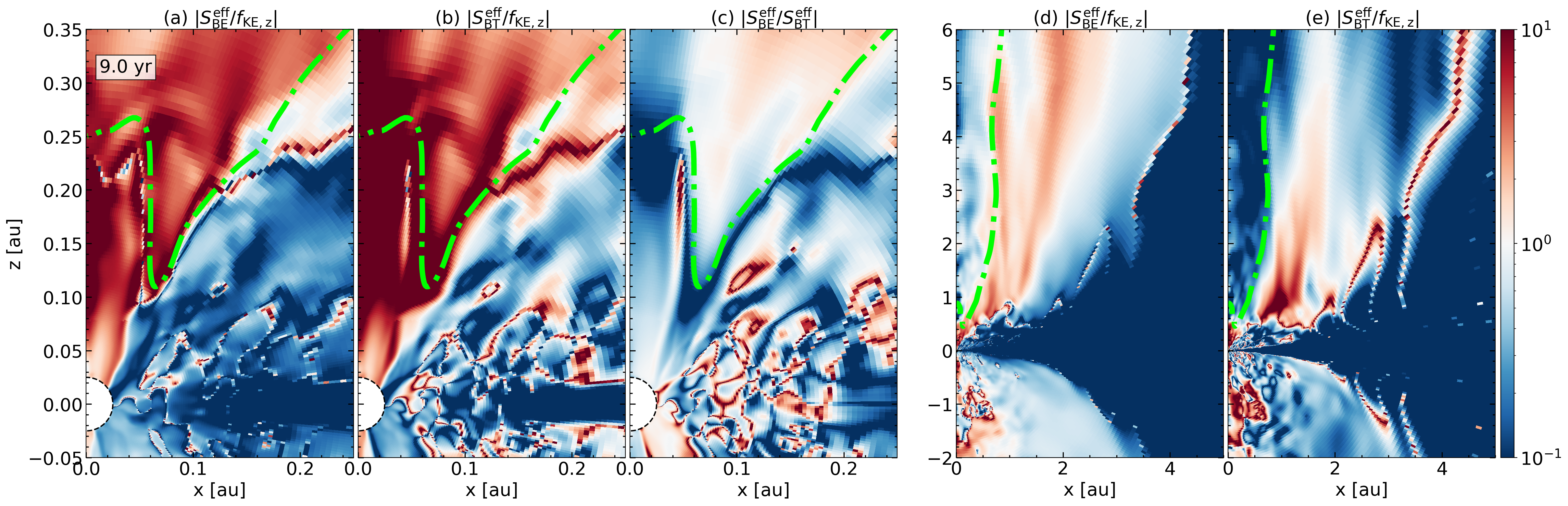}
    \caption{Ratios between energy advection through two components of the Poynting flux and gas kinetic energy {\it for the jet}, outlined by the cyan contour. The left three panels focus on the base of the jet, and the cyan contour is at $10^6~\mathrm{cm\ s^{-1}}$. The right two panels zoom out to larger scales, with the cyan contour at $10^7~\mathrm{cm\ s^{-1}}$. \textbf{Panel (a)} shows the ratio between the magnetic pressure component of the Poynting flux (equ.~\ref{equ:SBE_eff}) and kinetic energy flux in the $\hat{z}$ direction; \textbf{panel (b)} shows the ratio between the magnetic tension component of the Poynting flux (equ.~\ref{equ:SBT_eff}) and kinetic energy flux, and \text{panel (c)} shows the ratio between magnetic energy and tension components of the Poynting flux, illustrating the dominance of energy transportation by the advection of the toroidal field. \textbf{Panels (d) and (e)} show the same ratios as panels (a) and (b), but on a larger scale. Kinetic energy flux gradually becomes more important as the jet propagates to a larger distance. }
    \label{fig:poyntingflux}
\end{figure*}

The low-density polar cavity above the avalanche accretion stream is an ideal environment for jet launching. The magnetic field generated in the disk and along the avalanche accretion stream stores a large amount of energy (Fig.~\ref{fig:r18_azimavg_props}[c]), and it can drive a jet if the energy is transferred to a small amount of mass. To examine the jet-launching mechanism from an energetic perspective, we demonstrate energy advection and conversion in the jet-launching cavity in this section.

The energy advection is described by the MHD energy equation
\begin{equation}
    \frac{dE}{dt} + \nabla\cdot\Big[\Big(\frac{1}{2}\rho \mathbfit{v}^2 + P + h + \frac{B^2}{4\pi}\Big) \mathbfit{v} - \frac{1}{4\pi}\mathbfit{B}(\mathbfit{B}\cdot\mathbfit{v}) + \mathbfit{Q}\Big] = 0,
\end{equation}
where $P$ is the gas thermal pressure; $h$ is the gas internal energy and $\mathbfit{Q}$ is the heat flux. Since magnetic and fluid kinetic energy fluxes dominate in a jet-launching system, we can keep only two terms while ignoring the others: the kinetic energy flux
\begin{equation}
    f_\mathrm{KE} = \frac{1}{2}\rho v^2\mathbfit{v}
\end{equation}
and magnetic energy flux, which is equivalent to the Poynting flux
\begin{equation}
    \mathbfit{S} = \frac{1}{4\pi}(-\mathbfit{v}\times \mathbfit{B})\times \mathbfit{B} = \frac{1}{4\pi}\Big[B^2\mathbfit{v} - \mathbfit{B}(\mathbfit{B}\cdot\mathbfit{v})\Big]. 
    \label{equ:Poynting_flux}
\end{equation}
The Poynting flux consists of a magnetic energy advection term and a magnetic tension energy term. In the jet-launching cavity, we focus on the energy advection in the $\hat{z}$ direction. Using a similar argument as in decomposing the magnetic force into effective pressure and effective tension by removing the two terms canceling each other (as done in the derivation of equ.~\ref{equ:az_mag_pres_eff}, \ref{equ:az_mag_tens_eff}), we can define the ``effective magnetic energy advection'' as 
\begin{equation}
    S_\mathrm{BE}^\mathrm{eff} = \frac{1}{4\pi}(B_R^2 + B_\phi^2)v_z, 
    \label{equ:SBE_eff}
\end{equation}
and the ``effective energy advection due to magnetic tension'' as
\begin{equation}
    S_\mathrm{BT}^\mathrm{eff} = - \frac{1}{4\pi}(B_z B_R v_R + B_zB_\phi v_\phi).
    \label{equ:SBT_eff}
\end{equation}
To quantify how energy is advected in the jet-launching cavity, we show the azimuthally-averaged ratios between $S_\mathrm{BE}^\mathrm{eff}$, $S_\mathrm{BT}^\mathrm{eff}$, and $f_\mathrm{KE, z}$ in Fig.~\ref{fig:poyntingflux}. Panels (a), (b), and (c) focus on the small scale near the base of the jet; panels (d) and (e) show the former two ratios on larger scales. The cyan line outlines the rough location of the outflow. 
Below the base of the outflow (below the dashed cyan line in fig.~\ref{fig:poyntingflux}), 
$S_\mathrm{BT}^\mathrm{eff}$ dominates energy transportation over both $S_\mathrm{BE}^\mathrm{eff}$ and $f_\mathrm{KE, z}$. Within the outflow, $S_\mathrm{BE}^\mathrm{eff}$ dominates over both $S_\mathrm{BT}^\mathrm{eff}$ and $f_\mathrm{KE, z}$: energy is transported mostly by the advection of toroidal magnetic field. At a larger scale, $f_\mathrm{KE, z}$ dominates both magnetic terms in the jet as the magnetic energy carried by the toroidal magnetic field is converted into kinetic energy. 

\begin{figure*}
    \centering
    \includegraphics[width=\linewidth]{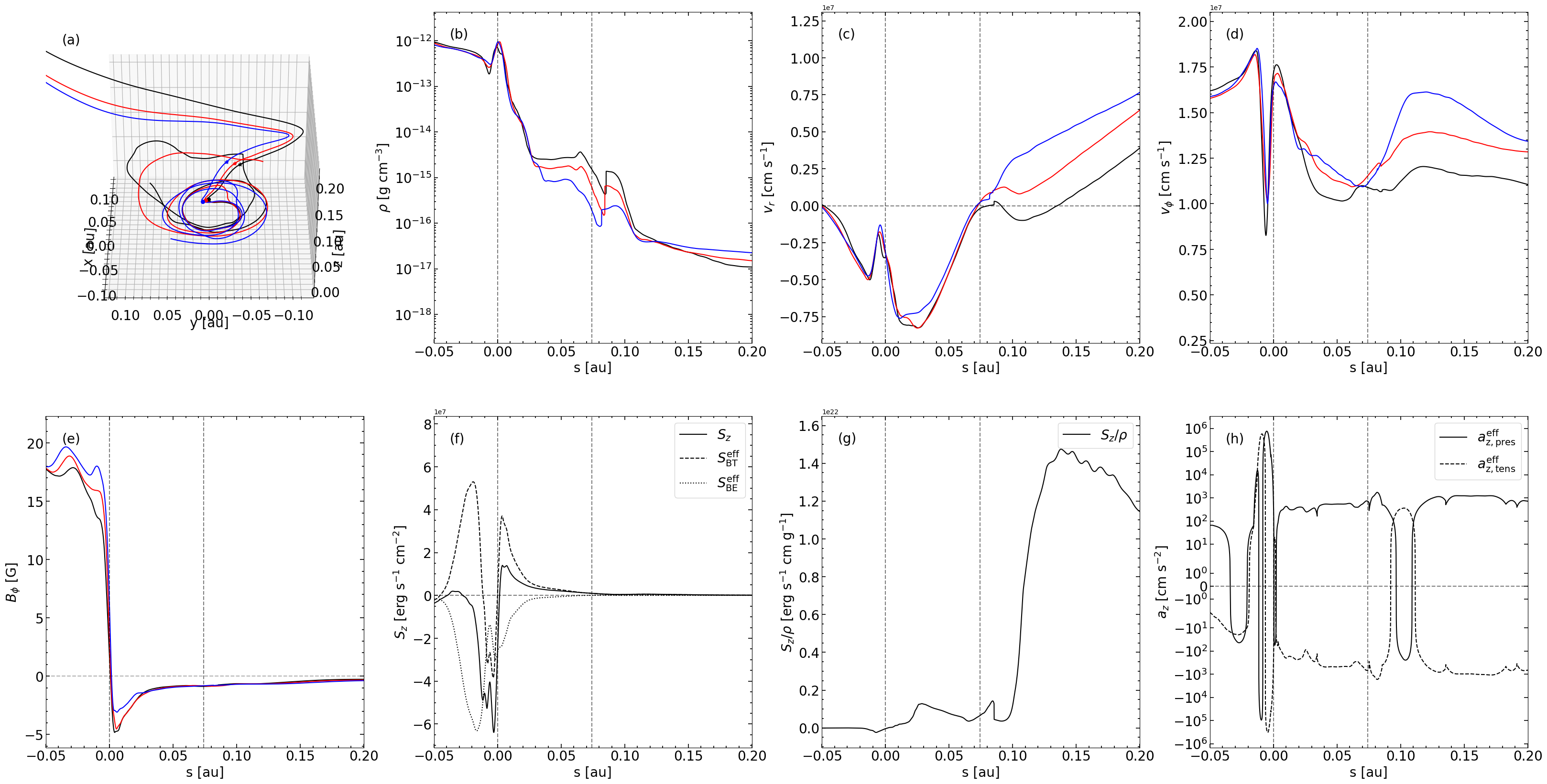}
    \caption{Quantities along magnetic field lines passing through the avalanche accretion stream. \textbf{Panel (a)} shows three consecutive fieldlines in 3D. The avalanche accretion stream is located at the pinching point of these fieldlines, marked by a solid dot. All other panels traces various quantities along these fieldlines, with $s = 0$ located at the solid dot and positive direction tracing the fieldline upward: the vertical dashed line at $s\approx 0.075$~au corresponds to the ``star'' along each magnetic fieldline. \textbf{Panels (b), (c), (d), and (e)} shows the density, radial direction velocity $v_r$, azimuthal velocity $v_\phi$, and toroidal magnetic field $B_\phi$ respectively, with line colors matching the respective field lines in panel (a). The similarities between these three magnetic field lines show a consistent evolution along the magnetic field lines advected by the avalanche accretion stream. \textbf{Panels (f), (g), and (h)} focuses on the black fieldline in panel (a), showing Poynting flux (and its terms, defined in equ.~\ref{equ:Poynting_flux}, \ref{equ:SBT_eff}, \ref{equ:SBE_eff} respectively), Poynting flux per unit mass, and magnetic acceleration on the gas (defined in equ.~\ref{equ:az_mag_pres_eff} and equ.~\ref{equ:az_mag_tens_eff}) respectively. 
    An animated version showing the 3D structure of the fieldlines is available at \url{https://figshare.com/s/ff9d053534f58e00f774}.
    The animation is 4 seconds long, showing a 3D view of panel (a)}
    
    \label{fig:fieldline_evolution}
\end{figure*}

To illustrate the connection between the avalanche accretion stream (Section~\ref{sec:avalanche accretion}) and energy release, Figure~\ref{fig:fieldline_evolution} traces the 3D spatial evolution of a representative jet-launching magnetic fieldline. Fig.~\ref{fig:fieldline_evolution}(a) depicts three consecutive fieldlines along the avalanche accretion stream, each passing through the magnetic pinching point (marked by dots) in the accretion stream, which serves as the reference origin ($s = 0$, the vertical dashed line in panels [b]-[g]) for analyzing physical quantities along the fieldlines. A second point on each fieldline marks the outflow base (where \(v_r = 0\)), corresponding to the other vertical dashed line (at $s\approx0.075$~au) in panels (b)-(g) to establish spatial context.

Fig.~\ref{fig:fieldline_evolution}(b) and~\ref{fig:fieldline_evolution}(c) reveal a stark density contrast (about five orders of magnitude) and radial velocity reversal between the dense accretion stream ($\sim 10^{-12}~\mathrm{g\ cm}^{-3}$) and the low-density outflow ($\sim 10^{-17}~\mathrm{g\ cm}^{-3}$). In Fig.~\ref{fig:fieldline_evolution}(d), the azimuthal velocity $v_\phi$ drops abruptly below the pinching point, signaling intense magnetic braking. The differential rotation around this braking point switches the sign of $B_\phi$ from positive in the disk to negative in the outflow, as shown in Fig.~\ref{fig:fieldline_evolution}(e). 

The similarities between the gas properties along the three consecutive fieldlines in Fig.~\ref{fig:fieldline_evolution} show a coherent evolution along the avalanche accretion stream where rapid infall, efficient magnetic braking, and magnetic field generation constantly supply twisted magnetic field to the outflow-launching cavity. In the following analysis, we will focus on one of the three highlighted fieldlines to illustrate the energy evolution. Immediately above the pinching point, the magnitude of (the negative) $B_\phi$ reaches a local maximum, and the energy is transported to the outflow-launching region primarily through magnetic tension ($S_\mathrm{BT}^\mathrm{eff}$): as the gas is infalling in this region, only magnetic tension can advect energy upstream into the outflow cavity. The magnetic energy flux per unit mass ($S/\rho$) is much higher in the polar cavity than in the disk (see panel [g]), making it possible to drive the jet to a high speed.

The dominance of magnetic tension in energy transport does not, however, imply its importance as the dominant force in accelerating the outflow. As shown in Fig.~\ref{fig:fieldline_evolution}(h), only the magnetic pressure force acts outward (away from the disk); the tension force remains directed toward the mid-plane--a result consistent with the geometric arguments in Section~\ref{sec:geometry}.

To summarize, the jet-launching mechanism in our model operates as follows: energy is transferred from the dense, rapidly rotating gas in the disk to the magnetic field through the winding of the field within the magnetically active inner disk zone. When a polar density cavity forms—particularly on the side of the disk where the toroidal magnetic field changes sign and creates the avalanche accretion stream—the energy stored in the toroidal field can be released into the gas within the low-density cavity, leading to the formation of a fast jet. The model with a rotating inner boundary, ``\textsc{strot}'', yields a qualitatively similar result, and we leave a comparison of the model ``\textsc{norot}'' without boundary rotation discussed here and the ``\textsc{strot}'' model to Appendix.~\ref{app:insensitivity}.

\section{Comparison to axisymmetric steady-state outflows}
\label{sec:compare_to_steady_state}

\begin{figure*}
    \centering
    \includegraphics[width=\textwidth]{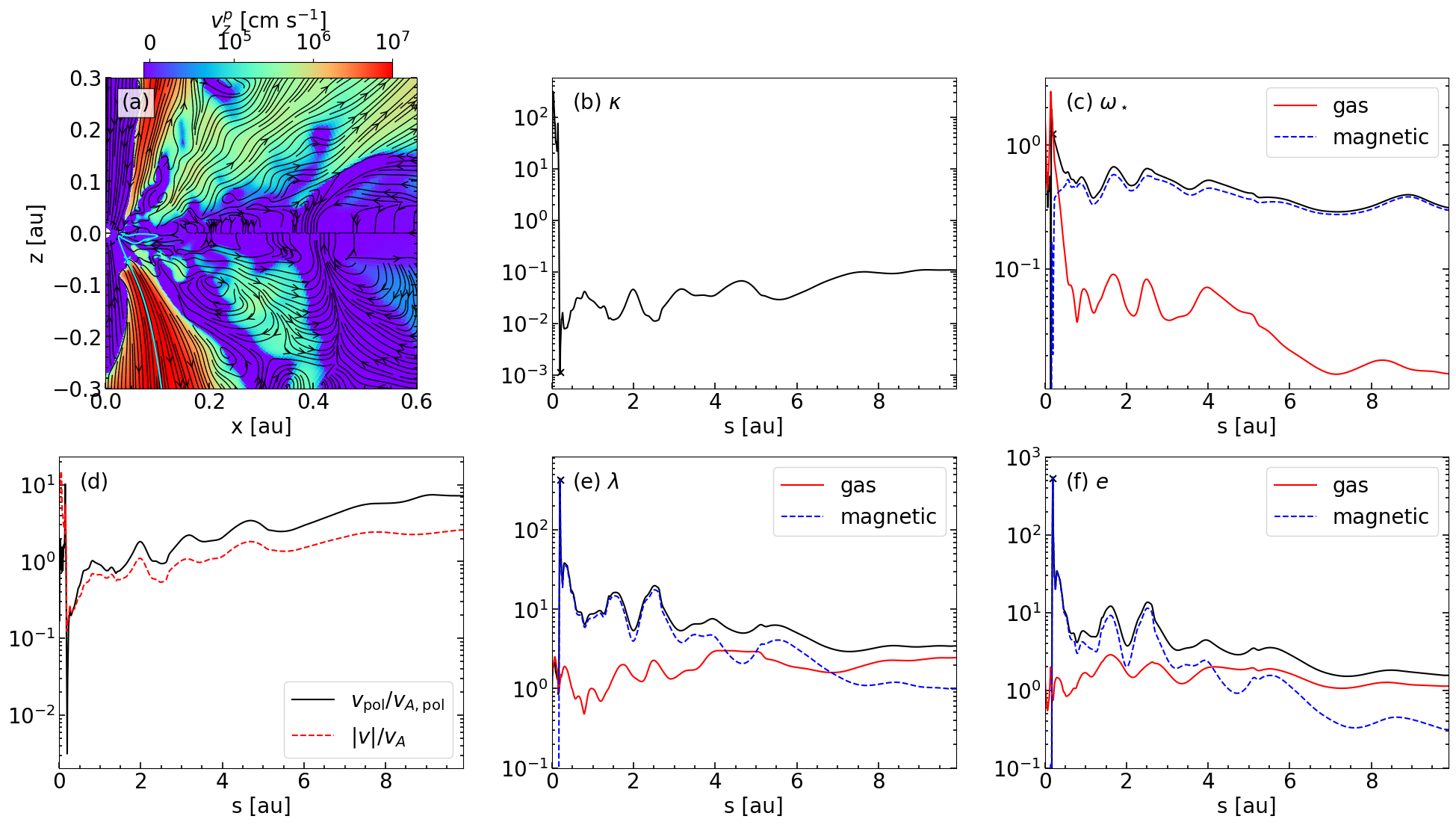}
    \caption{Quantities along a magnetic field line at the representative time $t=4.5$~yr. All values are azimuthally averaged to minimize numerical effects. \textbf{Panel (a)} shows the projected $v_z$ velocity (equ.~\ref{equ:vz-projection}). The representative field line analyzed is highlighted in cyan, where the positive direction traces the field line away from the disk midplane. \textbf{Panels (b), (c), (e), and (f)} shows the four dimensionless conserved quantity defined in equ.~\ref{equ:cons:k}, \ref{equ:cons:O}, \ref{equ:cons:l}, \ref{equ:cons:e} respectively. \textbf{Panel (d)} shows the Alfv$\acute{\text{e}}$nic Mach number and the poloidal 
    Alfv$\acute{\text{e}}$nic Mach number along the magnetic field line.}
    \label{fig:similar_tor_mc}
\end{figure*}

\subsection{Connection to the axisymmetric steady-state outflow}
\label{sec:similar_to_steady}


The jet launched in our model is highly variable in time and non-axisymmetric (as seen, e.g., in the animated version of Fig.~\ref{fig:jet3D} and Fig.~\ref{fig:r18_upperzslice_vzp}), but it shares some similarities with the axisymmetric steady-state MHD wind solutions. In this section, we will illustrate the similarities.  
 
In a steady state, an effective method for understanding the outflow is to analyze quantities along a magnetic field line. If the poloidal velocity aligns with the poloidal magnetic field, this approach is qualitatively analogous to examining gas flow along a velocity streamline. Fig.~\ref{fig:similar_tor_mc}(b), (c), (e), (f), and (h) are quantities along the magnetic field line highlighted by the cyan line in Fig.~\ref{fig:similar_tor_mc}(a). The starting point of the field line is in the disk on the disk midplane, and the positive direction traces backward (relative to the magnetic field direction) along the field line away from the midplane in the lower hemisphere. We chose the lower half of this fieldline because the jet is more prominent in the lower hemisphere at the plotted time of the simulation (4.5 yr). 

Kinematically, the gas changes from sub-Alfv$\acute{\text{e}}$nic to super-Alfv$\acute{\text{e}}$nic along a magnetic field line as a result of the conversion from magnetic energy to kinetic energy. Fig.~\ref{fig:similar_tor_mc}(d) shows the poloidal Alfv$\acute{\text{e}}$n Mach number, calculated using only the poloidal component of the gas velocity and magnetic field, and the Alfv$\acute{\text{e}}$n Mach number, calculated using all components of velocity and magnetic field. Because the outflow region is initially dominated by the magnetic field, the gas first becomes super-poloidal Alfv$\acute{\text{e}}$nic at $\sim 2$~au. At $\sim 4$~au, the gas is fully super Alfv$\acute{\text{e}}$nic, and the importance of the magnetic field decreases along the magnetic field line.

In a steady-state axisymmetric MHD wind solution, there are four conserved quantities along a magnetic field line.  Although the outflow in our model is highly time-dependent, it is still instructive to analyze these conserved quantities to make a connection between the toroidal-field launched jet model advocated in this paper and the classic magneto-centrifugal model \citep[][]{Salmeron2011, Zhu2018}. The first of the four quantities is
\begin{equation}
    k = \frac{4\pi\rho v_\mathrm{pol}}{B_\mathrm{pol}},
    \label{equ:cons_mass}
\end{equation}
which corresponds to the mass loading of the outflow. The second is
\begin{equation}
    \omega_s = \frac{v_\phi}{R} - \frac{k B_\phi}{4\pi \rho R}.
    \label{equ:cons_omega}
\end{equation}
The third quantity
\begin{equation}
    l = R v_\phi - \frac{RB_\phi}{k}
    \label{equ:cons_angmom}
\end{equation}
corresponds to the angular momentum transported from the disk to the wind \citep[][]{Wang2019}. The last is
\begin{equation}
    E = \frac{1}{2}v^2 + \Phi - \frac{\omega_s R B_\phi}{k} + h,
    \label{equ:cons_ene}
\end{equation}
the Bernoulli invariant corresponding to energy flux conservation along a field line \citep[][]{Shu2000, Konigl2000}. $h$ is the thermal energy, which is negligible compared to the magnetic energy and kinetic energy (Fig.~\ref{fig:r18_azimavg_props},\ref{fig:r18_azimavg_outer_vz_betas_l}). Following \citet{Tu2024c} and \citet{Jacquemin-Ide2021}, we can non-dimensionalize these quantities using the gas quantities at the slow magneto-sonic point (denoted by SM, the location of the slow magneto-sonic point is marked by the "X" in Fig.~\ref{fig:similar_tor_mc})
\begin{equation}
    \kappa = k \frac{v_\mathrm{K, SM}}{B_\mathrm{z, SM}}
    \label{equ:cons:k}
\end{equation}
\begin{equation}
    \omega_\star = \frac{\omega_s}{\Omega_\mathrm{K, SM}};
    \label{equ:cons:O}
\end{equation}
\begin{equation}
    \lambda = \frac{l}{R_\mathrm{SM} v_\mathrm{K, SM}};
    \label{equ:cons:l}
\end{equation}
\begin{equation}
    e = \frac{E}{v^2_\mathrm{K,SM}},
    \label{equ:cons:e}
\end{equation}
where the subscript $K$ denotes the Keplerian value. 

Fig.~\ref{fig:similar_tor_mc} shows these four dimensionless conserved quantities along the magnetic field line highlighted in Fig.~\ref{fig:similar_tor_mc}(a). The mass loading parameter decreases from $10^2$ to $10^{-2}$ as the field line goes from inside the disk into the jet-launching region and stays below $10^{-1}$ along the outflow, indicating that the magnetic field in the jet is lightly loaded with mass, as in the reference magneto-centrifugal wind solution of  \citet{Blandford1982}. However, the normalized angular velocity, $\omega_\star$, while staying roughly constant along the field line, is substantially below unity (with an average value of order $\sim 0.5$), unlike the standard lightly loaded magneto-centrifugal wind. 
The angular momentum $\lambda$ and the energy density $e$ both hold almost a constant along the fieldline, despite a slight decrease at large distances. They are dominated by the magnetic field at small radii and transition to being gas-dominated along the fieldline at high altitudes ($\sim 5$~au). These findings are broadly consistent with the results of \citet{Tu2024c}.

\subsection{Differences from the magneto-centrifugal model}
\label{sec:comp_magneto-centrifugal}

The magneto-centrifugal model \citep[as reviewed by, e.g.,][and references within]{Spruit2010} proposes that jets are launched by centrifugal forces, as angular momentum is transferred outward along rigid magnetic field lines anchored in a rotating system. In this subsection, we illustrate that, despite some similarities between our model and the magneto-centrifugal model, the requirements and expectations of the magneto-centrifugal mechanism are not met in our model. 

\begin{figure}
    \centering
    \includegraphics[width=\linewidth]{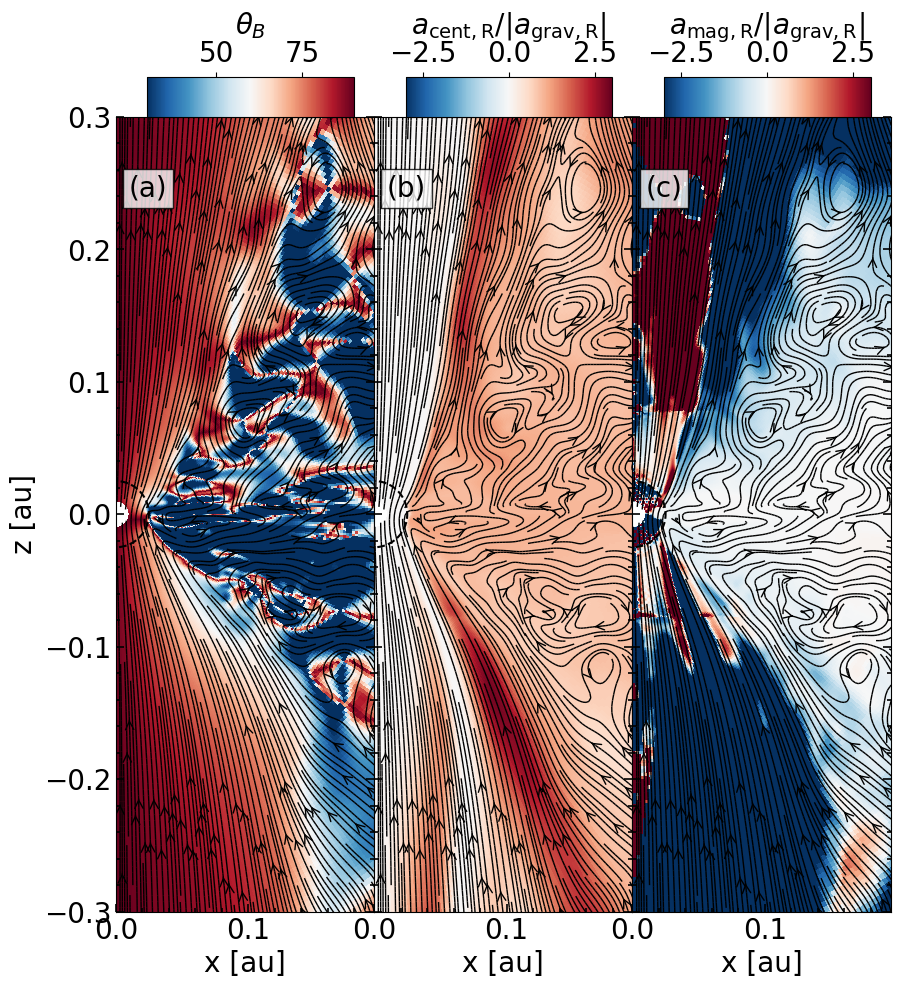}
    \caption{Deviation from the expectations of the magneto-centrifugal model. \textbf{Panel (a)} shows the pitch angle of the poloidal magnetic field line is more than 60 degrees from the disk midplane; \textbf{panel (b) and (c)} shows although centrifugal force is strong in the jet, magnetic force in the cylindrical-$\hat{R}$ direction is stronger. }
    \label{fig:diff_magneto_centrifugal}
\end{figure}

Geometrically, the poloidal magnetic field line must be inclined $<60$ degrees from the disk midplane for the centrifugal force to be effective in initiating a cold outflow from a thin disk \citep[e.g.][]{Krasnopolsky1999}, i.e.
\begin{equation}
    \theta_B = \arctan\Big|\frac{z}{R}\Big|,
\end{equation}
where $R$ is the cylindrical radius. Fig.~\ref{fig:diff_magneto_centrifugal}(a) shows that the poloidal magnetic field in our jet inclines by $\theta_B> 60$ degrees from the disk midplane, deviating from the expectations in the standard magneto-centrifugal model. However, the jet has a higher specific angular momentum compared to that at their launching location due to the transport of angular momentum by magnetic field lines (Fig.~\ref{fig:r18_azimavg_props}). As a result, the centrifugal force is stronger than the gravity in the cylindrically radial direction, although it is still weaker than the magnetic force that collimates the jet towards the axis (see Fig.~\ref{fig:diff_magneto_centrifugal}[b] and [c]).  

\subsection{Differences from axisymmetric steady-state jets}
\label{sec:comp_steady-state}

As mentioned in sec.~\ref{sec:similar_to_steady}, in an axisymmetric steady-state jet model, there are four conserved quantities (equ.~\ref{equ:cons_mass}, \ref{equ:cons_omega}, \ref{equ:cons_angmom}, \ref{equ:cons_ene}) along magnetic field lines. In such a model, we can establish a relationship between the specific angular momentum $Rv_\phi$, the outflow poloidal velocity $v_\mathrm{pol}$ at large distances from its base, and the radius of the outflow launching point, assuming the outflow originates from the disk plane where specific angular momentum is dominated by gas angular momentum, i.e.
\begin{equation}
    l = l_0 = R_0 v_{\phi, 0}, 
\end{equation}
\begin{equation}
    \omega_s = \frac{v_{\phi, 0}}{R_0} = \Omega_0.
    \label{equ:Ferreira2006_omega}
\end{equation}
where the subscript ``$0$" denotes the footpoint of the outflow on the disk plane. The angular velocity $\omega_s$ can be substituted into conserved energy:
\begin{equation}
    E = \frac{1}{2}v_{\phi}^2 - \Phi - \frac{\Omega_0 R B_\phi}{k} + h.
    \label{equ:Ferreira2006_ene}
\end{equation}
Following \citet[][]{Ferreira2006}, we can use equ.~\ref{equ:Ferreira2006_ene} to parameterize $R v_\phi$ and $v_\mathrm{pol}$ into
\begin{equation}
    Rv_\phi = \sqrt{GMR_0}\delta_0\lambda_\phi
    \label{equ:Ferreira2006_Rvphi}
\end{equation}
\begin{equation}
    v_\mathrm{pol} = \sqrt{\frac{GM}{R_0}}\sqrt{\delta_0^2(2\lambda_p-1)-2+\beta}
    \label{equ:Ferreira2006_vpol}
\end{equation}
where
\begin{equation}
    \delta_0^2 = \frac{\Omega_0^2R_0^3}{GM};
\end{equation}
\begin{equation}
    \lambda_\phi = (1-g)\frac{R^2}{R_0^2};
    \label{equ:lambda_phi}
\end{equation}
\begin{equation}
    \lambda_p = \lambda_\phi\frac{1+g}{2};
\end{equation}
\begin{equation}
    g = 1 - \frac{\Omega}{\Omega_0};
\end{equation}
\begin{equation}
    \beta = 2\frac{h_0 - h}{GM/R_0},
\end{equation}
assuming no heat input along the outflow. $\Omega$ is the local gas angular velocity. \citet[][]{Ferreira2006} divides the solutions to $Rv_\phi$ and $v_\mathrm{pol}$ into two regimes: the ``disk wind'' regime, where thermal energy is assumed to be unimportant ($\beta = 0$) and $\lambda_p\approx\lambda_\phi$, and the ``stellar wind'' regime, where thermal pressure $\beta$ dominates poloidal velocity. 

\begin{figure*}
    \centering
    \includegraphics[width=\textwidth]{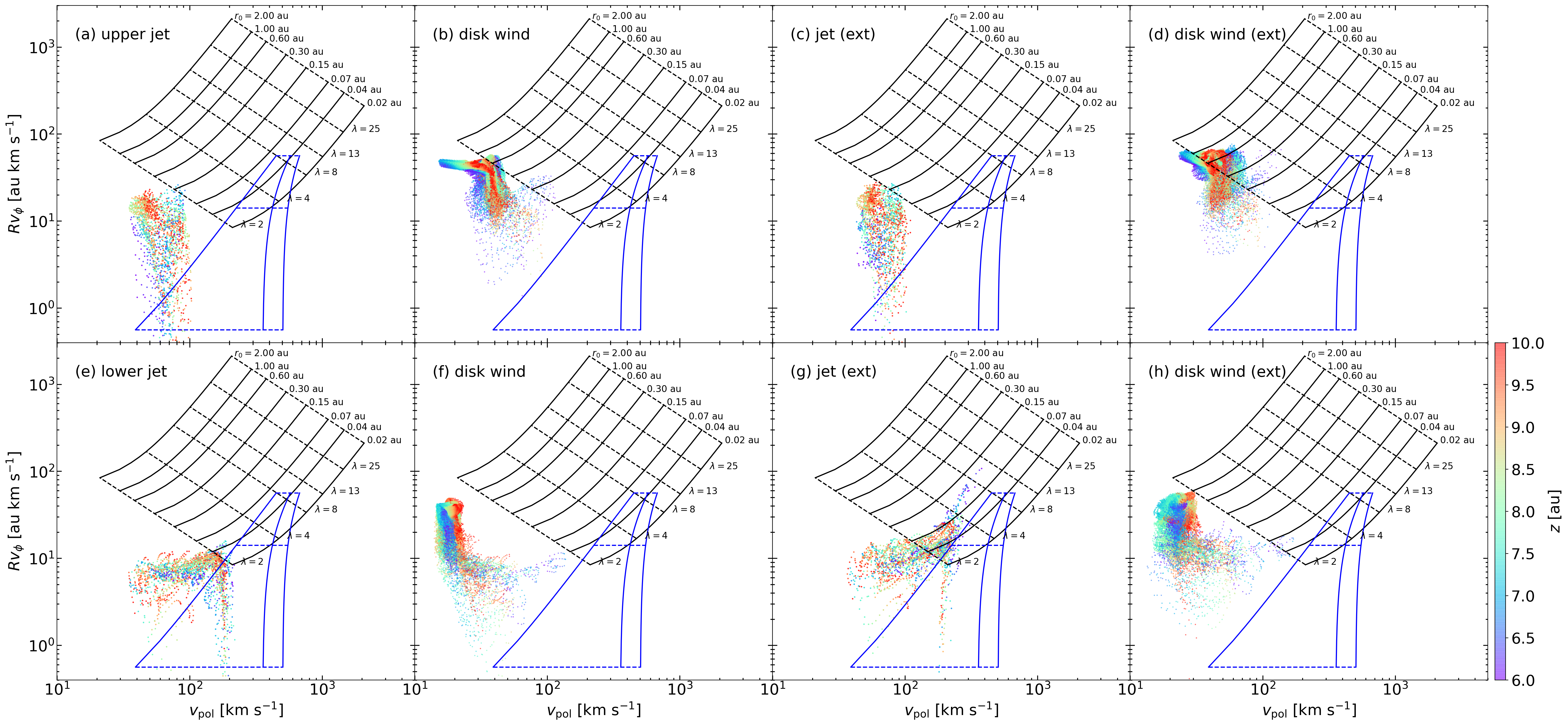}
    \caption{Comparison between the jet and disk wind in our model and the steady-state solutions from \citet[][]{Ferreira2006}. The outflow is divided into the jet and disk wind based on geometry (see sec.~\ref{sec:comp_steady-state}). Each cell is represented by a dot, colored according to its vertical distance from the disk midplane ($|z|$). \textbf{Panels (a) and (b)} show the jet and disk wind in the upper hemisphere, respectively; \textbf{panels (c) and (d)} show the same cells as a and b, but with values extrapolated assuming all angular momentum and kinetic energy carried by the magnetic field are released into the gas. The lower panels present the same quantities as the upper panels but use cells in the lower hemisphere.}
    \label{fig:Rvphi_vpol}
\end{figure*}

Fig.~\ref{fig:Rvphi_vpol} shows the ``disk wind'' and ``stellar wind'' regimes adapted from \citet[][]{Ferreira2006}. To compare our model to the steady-state solutions, we select only the cells greater than $6$ au from the disk midplane ($|z| > 6~$au) and $v_z^p > \max(10^6~\mathrm{cm\ s^{-1}}, v_\mathrm{escape})$, where $v_\mathrm{escape}$ is the local escape velocity. Since the outflow is asymmetric about the midplane, we show the selected upper hemisphere in the upper row of Fig.~\ref{fig:Rvphi_vpol} and the lower hemisphere in the lower row of Fig.~\ref{fig:Rvphi_vpol}. Because the outflow in our model can be divided into the jet and the disk wind (sec.~\ref{sec:result}), we divide our simulation into two regimes by geometry: the jet, which includes gas within $0.1 |z|$ from the polar axis, and the disk wind, which includes the outflow up to $4~\mathrm{au}$ in the cylindrical-$\hat{R}$ direction. Each selected cell is represented by a dot in Fig.~\ref{fig:Rvphi_vpol}, color-coded by the distance between the cell and the disk midplane (i.e., $|z|$). In both hemispheres, the jet resides beyond the lower left corner of the disk-wind parameter space and exhibits a wide vertical distribution. The disk wind is also clustered off the disk wind parameter space, moving slower but carrying more angular momentum than the jet.

The primary difference between the simulation and the steady-state solutions arises from the assumption regarding the angular momentum at the base of the outflow (equ.~\ref{equ:Ferreira2006_omega}). In our model, due to the highly dynamic and time-dependent nature of the disk's magnetically active zone at the outflow's base, conserved quantities are only applicable after the outflow is launched. For instance, as shown in Fig.~\ref{fig:similar_tor_mc}, these quantities are conserved only after some distance beyond the slow magnetosonic point along a magnetic field line. If we position 
the ``footpoint'' of the outflow somewhere near the slow magnetosonic point—outside the disk but within the outflow region, the conserved quantities are dominated by the terms with the toroidal field, making it necessary to account for the magnetic component of the conserved angular velocity at the footpoint (denoted again by the subscript ``0"), i.e.
\begin{equation}
    \omega_{s, 0} = \frac{v_{\phi, 0}}{R_0} + \frac{k B_{\phi, 0}}{4\pi\rho_0 R_0}
\end{equation}
and equ.~\ref{equ:Ferreira2006_ene} can be updated accordingly
\begin{equation}
    E = \frac{1}{2}v_{\phi}^2 - \Phi - \frac{\Omega_0 R B_\phi}{k} + h + \frac{B_\phi B_{\phi, 0} R}{4\pi\rho_0 R_0},
    \label{equ:Ferreira2006_ene_modified}
\end{equation}
which includes an extra term (the last term on the right hand side). At the footpoint of the outflow (where $R=R_0$), this term is twice the toroidal magnetic energy density $E_{B, \phi, 0} = \frac{B_{\phi, 0}^2}{4\pi\rho_0}$; it represents the Poynting flux, which is dominated by the $B_{\phi, 0}^2/(4\pi\rho_0)$ term (sec.~\ref{sec:jet_as_energy_release}). Far away from 
the footpoint, specific angular momentum is dominated by the gas component ($l$ in Fig.~\ref{fig:similar_tor_mc}), so $RB_\phi \ll R_0B_{\phi, 0}$. Defining
\begin{equation}
    \chi = 2\frac{\frac{B_{\phi, 0}}{4\pi\rho_0}(R_0B_{\phi, 0} - RB_\phi)}{GM/R_0} \approx 2\frac{E_{B, \phi, 0}}{GM/R_0},
\end{equation}
we can rewrite the poloidal velocity (equ.~\ref{equ:Ferreira2006_vpol}) as
\begin{equation}
    v_\mathrm{pol} = \sqrt{\frac{GM}{R_0}}\sqrt{\delta_0^2(2\lambda_p - 1) - 2 + \beta + \chi}.
\end{equation}
If $\chi$ is much larger than $\delta_0^2$, 2, and $\beta$ (i.e., if the toroidal magnetic energy flux at the footpoint of the jet is much larger than any other energy flux along the magnetic field line), then $v_\mathrm{pol}$ only depends on $\chi$. However, because $Rv_\phi$ is $\chi$-independent, there is no correlation between $v_\mathrm{pol}$ and $Rv_\phi$. This is the case for the jet in our model: the vertical distribution in Fig.~\ref{fig:Rvphi_vpol}(a) and (e) shows the lack of correlation between $v_\mathrm{pol}$ and $Rv_\phi$. As a consequence, measuring $v_\mathrm{pol}$ and $Rv_\phi$ does not directly yield the launching radius of the jet produced in our simulation. This is not surprising in hindsight because our jet is launched by the twisting of the magnetic field anchored on highly dynamic, fast-infalling, elevated avalanche accretion streams, which is very different from the conditions envisioned in the steady-state MHD wind launched from the surface of a slowly accreting, geometrically thin disk \footnote{ We should note that there are other effects that can change the relation between $v_\mathrm{pol}$ and $Rv_\phi$ for outflows in general, independent of the driving mechanism, such as rapid flow deceleration, which can force part of the outflow to reverse its rotation direction \citep[][]{Sauty2012}.}.

If $\chi$ is comparable to, but does not dominate over, $\delta_0^2$, we can reinterpret $\chi$ by defining a characteristic radius $r_*$ such that $\omega_s = \Omega_*$, the angular velocity at $r_*$. The interpretation of equ.~\ref{equ:Ferreira2006_Rvphi} and \ref{equ:Ferreira2006_vpol} would not change assuming all values estimated for the footpoint of the outflow are at $r_*$ instead of $r_0$. Consequently, the estimated outflow-launching radius based on equ.\ref{equ:Ferreira2006_Rvphi} and \ref{equ:Ferreira2006_vpol} will be $r_*$, which is less than the actual launching radius $r_0$. To connect our model with this interpretation, we assume at a large distance, all conserved quantities are dominated by the gas component, so the specific angular momentum is given by the conserved quantity $l$, and the poloidal velocity is calculated assuming poloidal kinetic energy dominates total energy, i.e.
\begin{equation}
    v_\mathrm{pol} = \sqrt{2E}.
\end{equation}
The extrapolation is performed for all points in Fig.~\ref{fig:Rvphi_vpol}(a), (b), (e), and (f), and the result is shown in Fig.~\ref{fig:Rvphi_vpol}(c), (d), (g), and (h) for the jet and disk wind in the upper and lower hemispheres. Most extrapolated points move a little towards the upper right corner, but the overall distribution stays around a similar position as in the original values. The vertical distribution of the jet remains, again showing the lack of correlation between specific angular momentum and poloidal velocity. The disk wind in our model now overlaps a little with the ``disk wind regime'' of the steady-state solution, with an estimated launching radius $r_0 \approx 0.6~$au, showing a reasonable estimate when $\chi$ is not dominating.

\section{Discussion}
\label{sec:discussion}

\subsection{Mass and angular momentum fluxes carried by the jet and disk wind}
\label{sec:dis_angmom}

In sec.~\ref{sec:result}, we demonstrated that the outflow can be divided into two distinct components: the fast-moving jet and the slower disk wind surrounding the jet. Between these, we identified an MRI-active turbulent disk wind region, which lacks a coherent flow structure, effectively separating the jet from the disk wind. The MRI-active turbulent disk wind becomes a laminar disk wind at a higher altitude where MRI can no longer be sustained.

\begin{figure*}
    \centering
    \includegraphics[width=\linewidth]{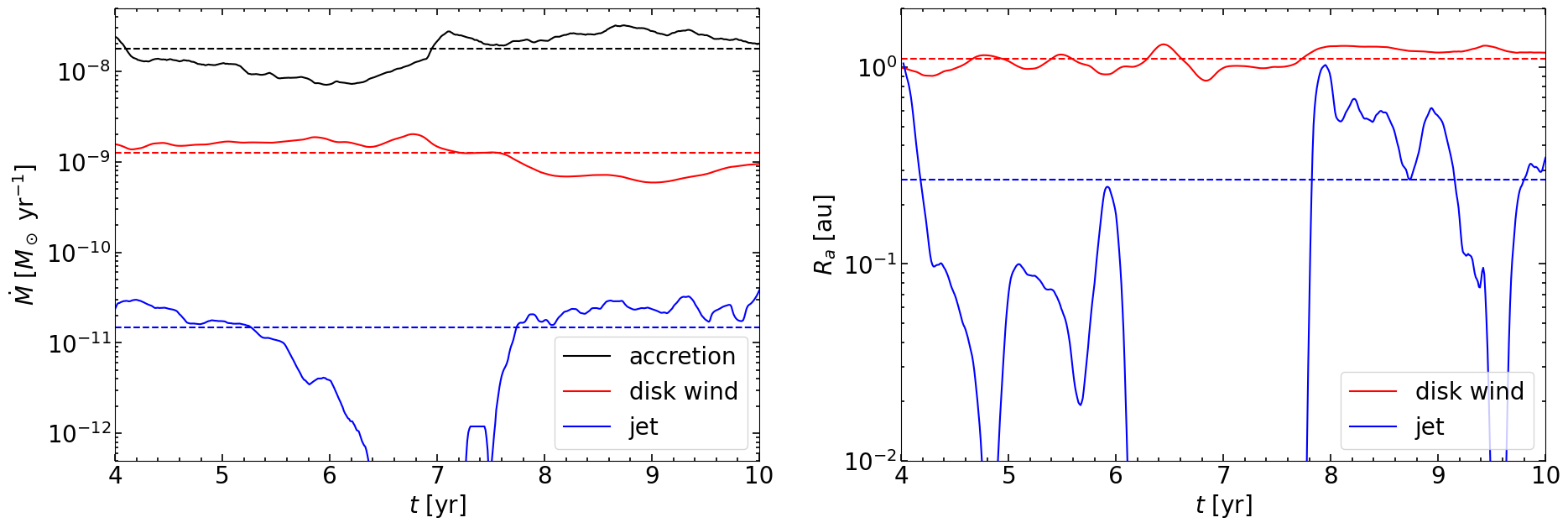}
    \caption{\textbf{Left panel: } Disk accretion rate and outflow mass flux. A negative sign is added to the accretion rate, so the values are positive. \textbf{Right panel: } The ``Keplerian angular momentum radius'' $R_a$ (defined in equ.~\ref{equ:keplerian_angmom_radius}) for the jet and the disk wind. The horizontal lines in both panels are the averaged values between 4.0 yr and 10.0 yr for each line. }
    \label{fig:mass_flux_and_Ra}
\end{figure*}

This outflow-launching picture naturally produces a well-collimated jet, where centrifugal force is not important compared with magnetic force, and a wider angle disk-wind, where centrifugal force acts to widen the opening angle of the outflow. We stress that in this picture, centrifugal force is not responsible for launching the jet or the disk wind: its primary role is to widen the opening angle of the disk wind. However, it does not translate to the absence of angular momentum transportation. Both the jet and disk wind are super-Keplerian (panel [d] in fig.~\ref{fig:r18_azimavg_props} and  \ref{fig:r18_azimavg_outer_vz_betas_l}). To quantify the mass and angular momentum advected by the outflow, we show in Fig.~\ref{fig:mass_flux_and_Ra}(a) the temporal evolution of the disk accretion rate and outflow mass flux at $z=\pm 10$~au, the outer simulation boundary in the vertical direction. The horizontal lines show the averaged values between 4.0 yr and 10.0 yr. The mass flux is defined as 
\begin{equation}
    \Phi_m = \int \rho \mathbfit{v}\cdot d\mathbfit{A},
\end{equation}
where $\mathbfit{A}$ is the area that the flow passes through. For accretion, $\mathbfit{A}$ is the sphere at 0.03~au, and for the outflow, $\mathbfit{A}$ is the planes at $z = \pm 10~\mathrm{au}$. A negative sign is added to the accretion mass flux, so all values are positive. The mass flux in the jet (gas moving at $v_z^p>10^7\ \mathrm{cm}\ \mathrm{s}^{-1}$) is about $2\times 10^{-11}\ M_\odot\ \mathrm{yr}^{-1}$, whereas the disk wind outflow mass flux ($v_z^p > 10^6\ \mathrm{cm}\ \mathrm{s}^{-1}$ and $R<4~\mathrm{au}$) is $10^{-9}\ M_\odot\ \mathrm{yr}^{-1}$; the latter is consistent with the empirical relation that the outflow-accretion mass flux ratio is about $1:10$ \citep[][]{Watson2016, Pascucci2023}. The relatively low mass flux in the fast jet may indicate that the magnetic field strength adopted in this paper in the inner disk may be too low to drive a fast jet with a much higher mass flux. For example, similar simulations but with a dynamically opened stellar magnetosphere that provides a much stronger field threading the inner disk can drive a fast jet with a much higher mass flux \citep[][]{Tu2025c}.

To quantify the angular momentum flux, we note that the momentum flux is defined as
\begin{equation}
    \Phi_l = \int \rho v_\phi R \mathbfit{v}\cdot d\mathbfit{A}.
\end{equation}
To connect the outflow momentum flux to the angular momentum on a Keplerian disk, we can define a ``Keplerian angular momentum radius'' $R_a$ such that
\begin{equation}
    v_K(R_a) R_a = \frac{\Phi_l}{\Phi_m},
\end{equation}
or, equivalently
\begin{equation}
    R_a = \frac{1}{GM}\Big(\frac{\Phi_l}{\Phi_m}\Big)^2.
    \label{equ:keplerian_angmom_radius}
\end{equation}
Fig.~\ref{fig:mass_flux_and_Ra}(b) shows the $R_a$ for the jet and the disk wind, respectively. The average $R_a$ for the jet is about $0.3$~au. Since the base of the jet is at $\lesssim 0.1$~au, the jet transports angular momentum per unit mass a few times higher than the specific angular momentum at their footpoint, consistent with the magnetic domination of the angular momentum near the base of the jet (see, e.g., Fig.~\ref{fig:similar_tor_mc}[e]). In contrast, the disk wind $R_a$ is about $1.0$~au, which is consistent with the interpretation of the steady-state solution (sec.~\ref{sec:comp_steady-state}) and its launching radii, which range from $\sim 0.1$~au to $\sim$~1.5 au, encompassing the disk wind $R_a$ of $1.0$~au.

\begin{figure}
    \centering
    \includegraphics[width=\linewidth]{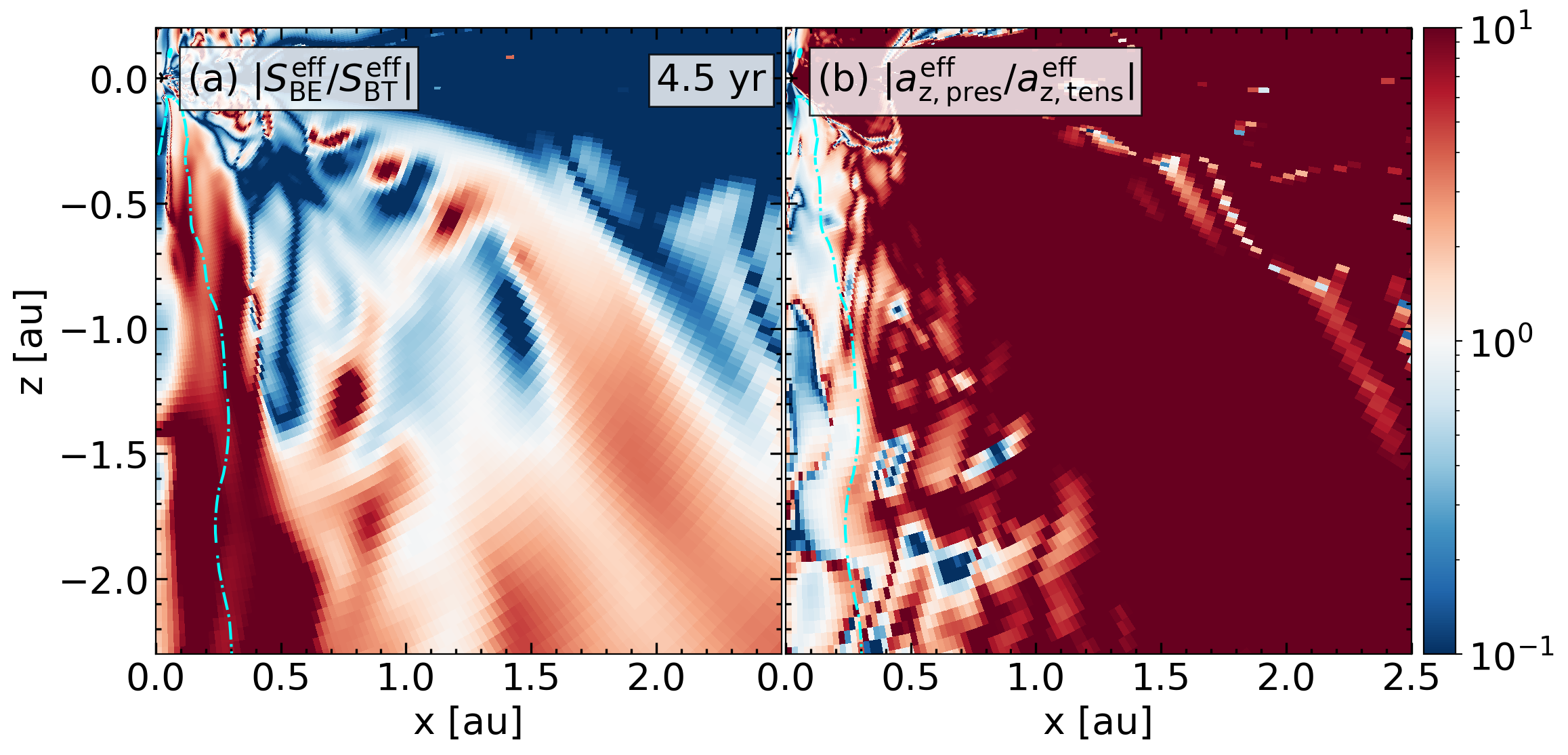}
    \caption{Ratios of energy advection rates and forces for the {\it disk wind}. The magnetic tension component of the Poynting flux (equ.~\ref{equ:SBT_eff}) dominates over the magnetic pressure component  (equ.~\ref{equ:SBE_eff}) in transporting energy in the $\hat{z}$ direction near the outflow base at small radii (left panel). However, the dominance in energy transport does not translate to a dominance in force: magnetic pressure force dominates over magnetic tension force everywhere in the disk wind (right panel).}
    \label{fig:Bernoulli_force}
\end{figure}

Angular momentum transport in the disk wind does not necessarily imply the magneto-centrifugal mechanism \citep[which is, in contrary to the usual interpretation in literature, e.g.,][]{Gressel2020}, which relies on the centrifugal force to launch an outflow. In particular, the dominance of the ``tension energy advection'' term (equ.~\ref{equ:SBT_eff}) over the ``pressure energy advection'' term (equ.~\ref{equ:SBE_eff}) of the Bernoulli invariant (sometimes used as direct evidence for the magneto-centrifugal mechanism) does not mean the magnetic tension force dominates over magnetic pressure force in the $\hat{z}$ direction. Fig.~\ref{fig:Bernoulli_force} shows a comparison between the energy flux ratio $S_\mathrm{BT}^\mathrm{eff}/S_\mathrm{BE}^\mathrm{eff}$ and the force ratios $a_\mathrm{z, tens}^\mathrm{eff}/a_\mathrm{z, pres}^\mathrm{eff}$ (defined in equ.~\ref{equ:az_mag_tens_eff} and equ.~\ref{equ:az_mag_pres_eff}), respectively. While it is true that the tension energy advection dominates over the pressure energy advection at the base of the disk wind, the magnetic pressure force overwhelmingly dominates the magnetic tension force. This contrast is not surprising as the kinetic-$\beta > 1$ in the disk wind region (Fig.~\ref{fig:r18_azimavg_outer_vz_betas_l}), 
so the tension energy advection, which includes the contribution from $v_\phi$, can appear significant without contributing to acceleration in the $\hat{z}$-direction. Moreover, the tension force itself points toward the disk midplane, making it inherently incapable of driving an outflow (see fig.~\ref{fig:r18_azimavg_forces} and sec.~\ref{sec:geometry}). The region where tension energy advection dominates is located at a small altitude. The flow transitions to magnetic pressure energy advection dominated at a larger altitude. The transition occurs when the gas switches from a predominantly horizontal motion along the cylindrical$\hat{R}$ direction to a predominantly vertical motion.

In contrast, the jet is toroidal magnetic pressure dominated in both force and energy advection (as it is overwhelmingly magnetically dominated, see Fig.~\ref{fig:r18_azimavg_props}), so the acceleration is primarily in the $\hat{z}$ direction and the opening angle of the jet stays small. We stress here again that efficient angular momentum transport is {\it not} evidence for the magneto-centrifugal mechanism. For the jet, angular momentum transport does not directly contribute to the flow acceleration (the two are related through the toroidal magnetic field generated by field line twisting near the jet base, however); for the disk wind, angular momentum in the wind increases the opening angle of the outflow through the centrifugal force.

\subsection{In-situ toroidal field generation}

We have shown the importance of the toroidal magnetic field in accelerating the jet and the disk wind. In this subsection, we demonstrate how the toroidal field is generated and advected in the jet-launching region.

\begin{figure*}
    \centering
    \includegraphics[width=\textwidth]{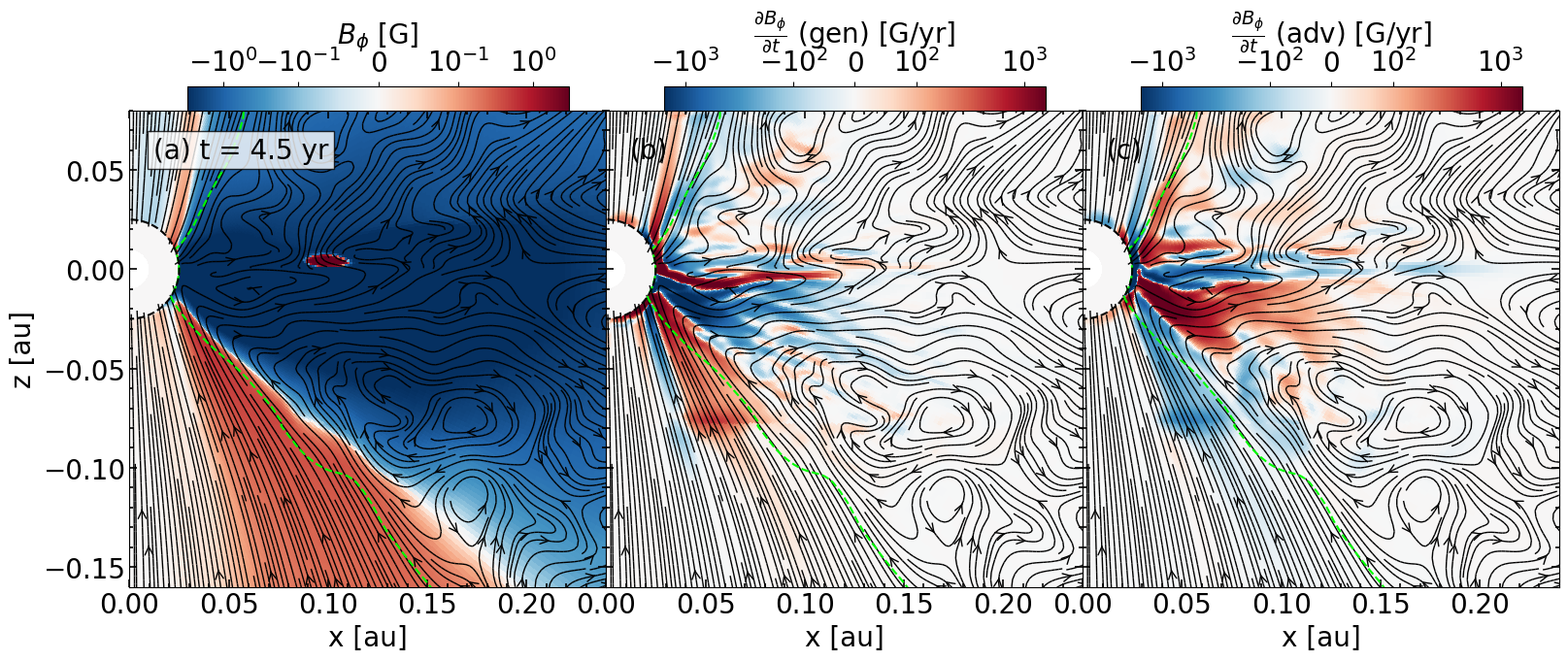}
    \caption{Azimuthally averaged magnetic field strength, generation, and advection. \textbf{Left panel:} azimuthal magnetic field $B_\phi$ at the representative time. \textbf{Middle panel}: generation of azimuthal magnetic field through differential rotation, as defined in equ.~\ref{equ:Bphi generation}; \textbf{Right panel:} advection of azimuthal magnetic field through gas motion, as defined in equ.~\ref{equ:Bphi advection}. The green line in each panel shows where $\beta_K = 1$ (defined in equ.~\ref{equ:kinetic-beta}), with the polar regions being magnetically dominated ($\beta_K < 1$). Note that a negative generation or advection rate in a negative $B_\phi$ region is an {\it{increase}} in magnetic field strength.}
    \label{fig:Bphi_gen_pbeta}
\end{figure*}

Toroidal magnetic generation is governed by the induction equation (equ.~\ref{equ:mhd induction}). In the magnetically active zone of the disk and the jet-launching region, the gas is in the ideal MHD limit, so we can ignore the second (Ohmic dissipation) term in equ.~\ref{equ:mhd induction}. The right-hand side of equ.~\ref{equ:mhd induction} in cylindrical coordinate in the azimuthal direction can be decomposed into two terms:
\begin{equation}
    \frac{\partial B_\phi}{\partial t}\Big|_\mathrm{gen} = 
    \frac{\partial (v_\phi B_z)}{\partial z} 
    + \frac{\partial (v_\phi B_R)}{\partial R}
    \label{equ:Bphi generation}
\end{equation}
\begin{equation}
    \frac{\partial B_\phi}{\partial t}\Big|_\mathrm{adv} = 
    - \frac{\partial (v_z B_\phi)}{\partial z} 
    - \frac{\partial (v_R B_\phi)}{\partial R}
    \label{equ:Bphi advection}
\end{equation}
where equ.~\ref{equ:Bphi generation} describes the generation of $B_\phi$ through differential twisting of the poloidal field and equ.~\ref{equ:Bphi advection} describes the advection of $B_\phi$ through poloidal gas motion. To show the evolution of the toroidal magnetic field, we show in Fig.~\ref{fig:Bphi_gen_pbeta} the toroidal magnetic field strength, and its generation and advection rates in panels (a), (b), and (c), respectively. As discussed in sec.~\ref{sec:jet_as_energy_release}, the jet in the lower hemisphere is accelerated in the $B_\phi > 0$ region. The generation of the jet-launching toroidal magnetic field occurs around the transition from $B_\phi < 0$ in the disk to $B_\phi > 0$ in the jet-launching cavity, where $\frac{\partial B_\phi}{\partial t}\Big|_\mathrm{gen} \approx 10^3\ \mathrm{G\ yr}^{-1}$, which, for a typical field strength of $\sim 1$~G in the region, corresponds to a very short timescale of only $\sim 10^{-3}$~yr!  The generated toroidal magnetic field is efficiently removed by advection at a comparable rate. A portion of this field is transported to the inner boundary via the avalanche accretion stream, which occurs where $B_\phi$ transitions from negative to positive (see Section \ref{sec:avalanche accretion}). The remaining portion of the field, situated outside the accretion stream, releases its energy into the low-density gas, propelling it in the (negative) $\hat{z}$-direction to form a toroidal magnetic pressure-driven jet in the lower hemisphere.

The rapid generation of the toroidal magnetic field is facilitated by the presence of a mass-dominated avalanche accretion stream. Within this stream, differential rotation twists the poloidal magnetic field, efficiently converting it into a toroidal magnetic field. The dominance by the kinetic energy is illustrated by the green line in Fig.~\ref{fig:Bphi_gen_pbeta}, which shows where $\beta_K = 1$(defined in equ.~\ref{equ:kinetic-beta}, see Fig.~\ref{fig:r18_azimavg_props}(c) for a 2D slice of $\beta_K$). The gas-dominated region, where $\beta_K > 1$, lies above the green line in the lower hemisphere and includes the fastest toroidal magnetic field generation region. On the other side of the green line, the toroidal field can still be generated {\it in situ} for a brief period within the magnetically-dominated region. 
%
This localized rapid toroidal field generation at the base of the polar cavity ensures a continuous supply of toroidal magnetic field, which provides the magnetic pressure (and the outward-going Poynting flux) needed to accelerate the low-density gas in the cavity to a high speed and form a fast jet. 

\section{Conclusions}
\label{sec:conclusion}

We performed a 3D non-ideal MHD simulation of outflow-launching from the inner region of a circumstellar disk threaded by open magnetic field lines. Our model includes a magnetically active disk zone surrounded by a magnetically dead zone. Although the jet in our model shares some similarities with the magneto-centrifugal model and the steady-state outflow solutions, the jet in our model is {\it not} driven magneto-centrifugally and has some key differences from the steady-state solutions. Our main conclusions are as follows
\begin{enumerate}
    \item The outflow can be divided into three components: (1) a jet--the collimated fast, lightly loaded outflow near the polar axis; (2) an MRI-active turbulent disk wind surrounding the jet where the flow is turbulent due to MRI at low latitudes but becomes laminar at high altitude where MRI can no longer be sustained; an (3) a disk wind, a laminar high-density slow outflow surrounding both the jet and the MRI-active turbulent disk wind.  It merges with the upper part of the MRI-active turbulent disk wind at high altitudes when the flow becomes laminar.
    
    \item The jet and the disk wind are both driven by toroidal magnetic pressure. A toroidal magnetic field is generated in the mass-dominated disk, where differential rotation twists the poloidal field into a toroidal field. The toroidal field generated at the surface of the disk releases its energy to the low-density gas in the polar cavity, producing the fast lightly loaded jet. Gas rotation plays a negligible role in driving the jet due to the dominance of magnetic field energy over both kinetic and thermal energies (both kinetic-$\beta$ and plasma-$\beta$ are $\ll 1$) in the jet acceleration region.

    \item In some extreme cases, the jet can be choked by the MRI-active turbulent disk wind when this wind fills the polar cavity with dense gas. The avalanche accretion stream is present only in one hemisphere due to the magnetic field geometry-facilitates jet launching by creating an interface between the turbulent disk wind and the polar cavity. This interface allows magnetic energy to be efficiently transferred to a small amount of gas, enabling jet acceleration. In the opposite hemisphere, where the avalanche stream is absent, the jet is more likely to be suppressed by the dense gas accumulated from the MRI-active wind. Nevertheless, the disk wind itself remains bipolar, as it is more heavily mass-loaded and does not rely on a polar cavity to be launched.
    
    \item The disk wind is also accelerated by the toroidal magnetic pressure, but unlike the jet, which is overwhelmingly magnetically dominated, the disk wind is kinetically dominated, so the centrifugal force can effectively widen the opening angle of the outflow.
    
    \item The jet in our model shares several similarities with the classic magneto-centrifugal model. Firstly, angular momentum is efficiently transported from the disk to the jet; second, the jet is lightly loaded; and thirdly, the jet transitions from sub-Alfv$\acute{\text{e}}$nic to super-Alfv$\acute{\text{e}}$nic along a magnetic field line. 
%
    %
    %
    However, the poloidal magnetic field lines in our jet are inclined by more than 60 degrees from the disk midplane, unlike the field configuration required for the initial flow acceleration from a geometrically thin disk in the standard lightly-loaded magneto-centrifugal wind. More importantly, even though our jet is lightly mass-loaded, it is primarily accelerated vertically by the unwinding of constantly twisted magnetic fields rather than horizontally by centrifugal force as ``beads on a wire" as envisioned in the classic magneto-centrifugal wind model.  
    
    \item The jet in our model also differs from the expectations of steady-state axisymmetric outflow solutions. Due to the highly variable and non-magneto-centrifugal nature of the outflow acceleration region near the jet base (where the jet is primarily powered by the unsteady magnetic field twisting from differential rotaton in high-altitude fast accretion streams), at a large distance where the flow is dominated by the gas, there is no correlation between the gas poloidal velocity and its specific angular momentum, so the poloidal velocity-angular momentum relation should not be used as an indicator of the jet-launching radius. This correlation can be used, with caution, to estimate the launching radius of the disk wind, as the disk wind is less magnetically dominated.

\end{enumerate}

\begin{acknowledgments}

YT acknowledges support from the interdisciplinary fellowship at UVA. CYH acknowledges support from NASA SOFIA grant SOF-10\_0505 and NRAO ALMA Student Observing Support (SOS). We acknowledge computing resources from UVA research computing (RIVANNA), NASA High-Performance Computing, and NSF's ACCESS computing allocation AST200032. ZYL is supported in part by NASA 80NSSC20K0533, NSF AST-2307199, and the Virginia Institute of Theoretical Astronomy (VITA). Z.Z. acknowledges support from NSF award 2408207, 2429732 and NASA grant 80NSSC24K1285.
\end{acknowledgments}

%






\appendix
\section{Insensitivity to the inner boundary condition}
\label{app:insensitivity}

In this Appendix, we demonstrate that our results are insensitive to the prescribed rotation at the inner boundary by comparing the \textsc{norot} model (where the inner boundary is stationary) and the \textsc{strot} model (where the inner boundary rotates as a solid body with a 2.9-day period). While these models are not expected to be identical due to their turbulent nature, we show that they are qualitatively similar.

\begin{figure*}
    \centering
    \includegraphics[width=\linewidth]{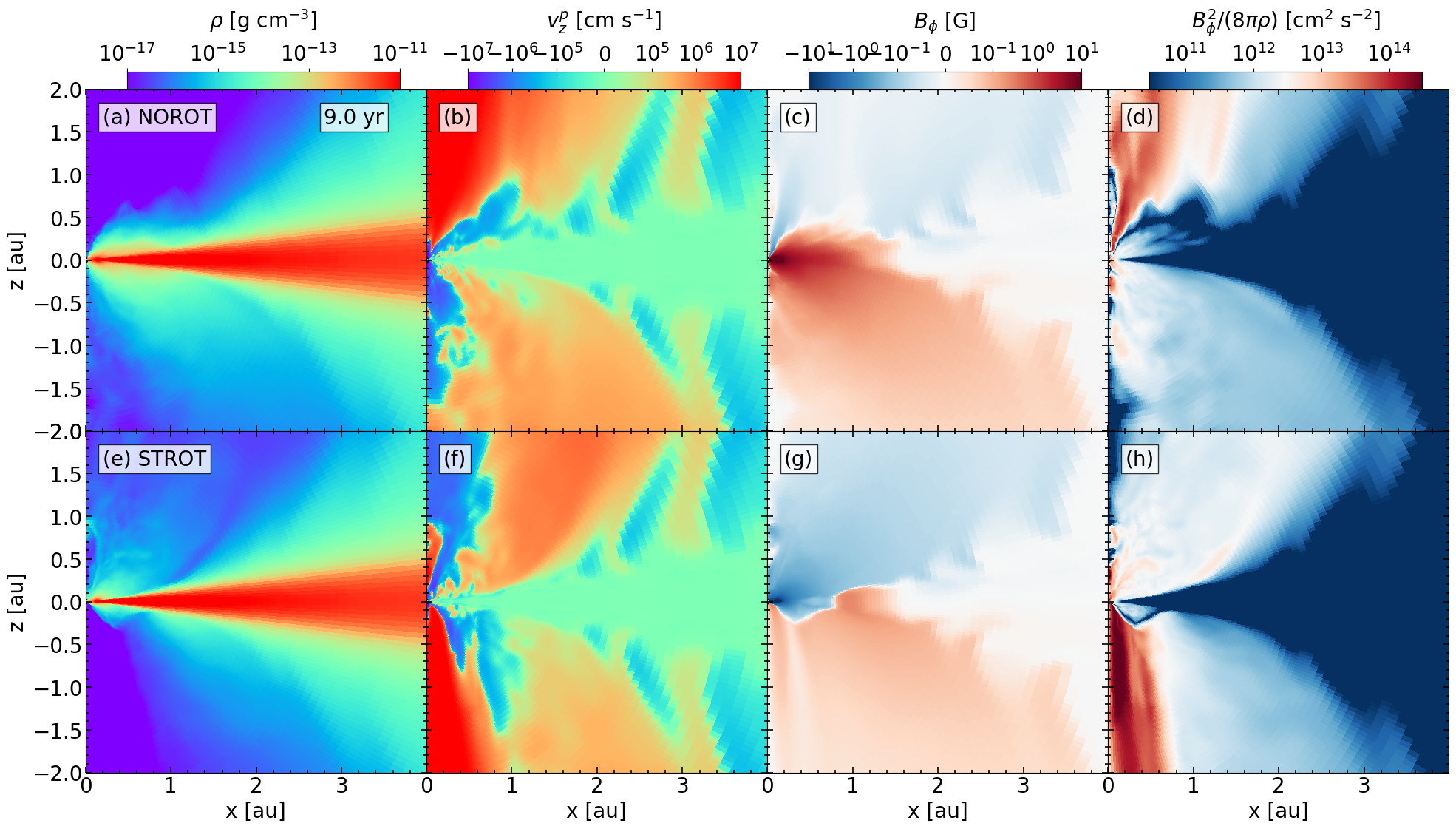}
    \caption{Comparison between the \textsc{norot} and \textsc{strot} at the same time (9.0 yr). The upper row shows a slice through the \textsc{norot} model and the lower row shows the \textsc{strot} model. The four columns show density, projected z-direction velocity ($v_z^p$, equ.~\ref{equ:vz-projection}), azimuthal magnetic field, and magnetic energy per unit mass, respectively. Although these two models are not expected to be identical due to their turbulent nature, they are qualitatively similar. An animated showing the evolution can be found at \url{https://figshare.com/s/04e8a31b7f92218f24bb}.
    The animated version is 50 seconds long, showing the evolution of the \textsc{norot} and \textsc{strot} models from the initial condition to 10 yrs. The animation shows the qualitative similarities between these two models.}
    \label{fig:comp_NOROT_STROT}
\end{figure*}

Figure~\ref{fig:comp_NOROT_STROT} presents the azimuthally averaged density, $v_z^p$, $B_\phi$, and $B_\phi^2/(8\pi\rho)$ for both the \textsc{norot} (upper row) and \textsc{strot} (lower row) models. Readers are encouraged to view the animated version for a clearer understanding of their similarities over time. In both models, the density structure features a magnetically expanded active zone near the midplane ($\lesssim 0.15$~au), a more laminar magnetic dead zone at larger radii ($\geq 0.15$~au), and an outflow extending above the disk in both hemispheres. At the time shown, the jet in the \textsc{norot} model is dominant in the upper hemisphere, while in the \textsc{strot} model, it is stronger in the lower hemisphere. The disk wind (sec.~\ref{sec:diskwind}) and MRI-active disk wind (sec.~\ref{sec:middle zone}) are prominent at large cylindrical radii in both hemispheres for both models.

The magnetic field configurations in the two models are consistent with the geometric arguments in Section~\ref{sec:geometry}: the azimuthal magnetic field $B_\phi$ is negative in the upper hemisphere and positive in the lower hemisphere. Although the hemisphere where $B_\phi$ switches sign differs—upper hemisphere in \textsc{norot} and lower hemisphere in \textsc{strot}—the jet-launching side is always aligned with the region where $B_\phi$ switches sign. Consequently, the magnetic energy per unit mass (fig.~\ref{fig:comp_NOROT_STROT}[d], [h]) is larger on the jet-launching side. This supports the conclusion that the disk surface avalanche accretion stream facilitates jet launching by creating an outflow cavity, a process that is robust and independent of the inner boundary's rotation.

The similarity between the two models might seem counterintuitive at first, as one might expect the additional differential rotation in the \textsc{strot} model to strengthen the outflow. However, in our simulations, the poloidal field lines threading the inner boundary do not exhibit significant differential rotation because the magnetic field dominates over gas motion immediately above and below the boundary. As the gas near the boundary adjusts to its motion, no additional differential rotation is introduced, causing the magnetic field lines in the polar regions to remain mostly vertical in both models. As a result, the \textsc{strot} model is qualitatively similar to the \textsc{norot} model. While the early jet in the \textsc{strot} model ($\lesssim 3$yr in the animated version of Fig.\ref{fig:comp_NOROT_STROT}) is indeed stronger, the two models converge to a qualitatively similar behavior in their later evolution.

\bibliography{sample631}{}
\bibliographystyle{aasjournal}



\end{document}